\documentclass[referee]{aa} % for a referee version

\usepackage{graphicx}
\usepackage{txfonts}
\usepackage{natbib}
\usepackage{eso-pic}
\usepackage{gensymb}
\usepackage{starfont}
\usepackage{diagbox}
\usepackage[normalem]{ulem}
\usepackage{paralist}
\usepackage{natbib,twoopt}
\usepackage[breaklinks=true]{hyperref}

\bibpunct{(}{)}{;}{a}{}{,}             %% natbib format for A&A and ApJ
\makeatletter
  \newcommandtwoopt{\citeads}[3][][]{\href{http://adsabs.harvard.edu/abs/#3}%
    {\def\hyper@linkstart##1##2{}%
     \let\hyper@linkend\@empty\citealp[#1][#2]{#3}}}
  \newcommandtwoopt{\citepads}[3][][]{\href{http://adsabs.harvard.edu/abs/#3}%
    {\def\hyper@linkstart##1##2{}%
     \let\hyper@linkend\@empty\citep[#1][#2]{#3}}}
  \newcommandtwoopt{\citetads}[3][][]{\href{http://adsabs.harvard.edu/abs/#3}%
    {\def\hyper@linkstart##1##2{}%
     \let\hyper@linkend\@empty\citet[#1][#2]{#3}}}
  \newcommandtwoopt{\citeyearads}[3][][]%
    {\href{http://adsabs.harvard.edu/abs/#3}
    {\def\hyper@linkstart##1##2{}%
     \let\hyper@linkend\@empty\citeyear[#1][#2]{#3}}}
\makeatother

\begin{document}

\title{Exoplanet Cartography using Convolutional Neural Networks}

\author{K.\ Meinke
          \inst{1,3}
          \and
          D.M.\ Stam
          \inst{1}
          \and
          P.M.\ Visser
          \inst{2}
          }

\institute{Faculty of Aerospace Engineering, Delft University of Technology, Kluyverweg 1, 2629 HS Delft, The Netherlands
    % \email{XXX}
    \and
    Delft Institute of Applied Mathematics, Delft University of Technology, Mekelweg 4, 2628 CD Delft, The Netherlands
    \and
    ORCID iD: \href{https://orcid.org/0000-0001-6696-1806}{0000-0001-6696-1806}}

\date{Received 16 December 2021; accepted 25 April 2022}

%%%%%%%%%%%%%%%%%%%%%%%%%%%%%%%%%%%%%%%%%%%%%%%%%%%%%%%%%%%%%%%%%%%%%%%%%%%%%%
\abstract
{In the near-future, dedicated telescopes will observe 
Earth-like exoplanets in reflected parent-starlight, allowing their physical characterization. Because of the huge distances, every exoplanet will remain an unresolved, single pixel, but temporal variations in the pixel’s spectral flux contain information about the planet’s surface and atmosphere.}
{We test convolutional neural networks for retrieving a planet’s rotation axis, surface and cloud map from simulated single-pixel observations of flux and polarization light curves. We investigate the influence of assuming that the reflection by the planets is Lambertian in the retrieval while in reality their reflection is bidirectional, and of including polarization in retrievals.}
{We simulate observations along a planet’s orbit using a radiative transfer algorithm that includes polarization and bidirectional reflection by vegetation, desert, oceans, water clouds, and Rayleigh scattering in six spectral bands from $400$ to $800~\text{nm}$, at various levels of photon noise. The surface-types and cloud patterns of the facets covering a model planet are based on probability distributions. Our networks are trained with simulated observations of millions of planets before retrieving maps of test planets.}
{The neural networks can constrain rotation axes with a mean squared error 
(MSE) as small as $0.0097$, depending on the orbital inclination. On a bidirectionally reflecting planet, $92\%$ of ocean facets and $85\%$ of vegetation, desert, and cloud facets are correctly retrieved, in the absence of noise. With realistic amounts of noise, it should still be possible to retrieve the main map features with a dedicated telescope. Except for face-on orbits, a network trained with Lambertian reflecting planets, yields significant retrieval errors when given observations of bidirectionally reflecting planets, in particular, brightness artefacts around a planet’s pole. Including polarization improves the retrieval of the rotation axis and the accuracy of the retrieval of ocean and cloudy map facets.}
{}

          
\maketitle

%%%%%%%%%%%%%%%%%%%%%%%%%%%%%%%%%%%%%%%%%%%%%%%%%%%%%%%%%%%%%%%%%%%%%%%%%%%%%%
\section{Introduction}

Since the discovery of the first exoplanet by \citet{first_exoplanet},
more than 4000 exoplanets have been identified and catalogued.\footnote{ \url{https://exoplanetarchive.ipac.caltech.edu/}}
Statistically, nearly every star has exoplanets
\citep{2020AJ....159..279B,occurence_of_earths,all_stars_have_planets_1},
and several of these will be rocky and in the habitable zone of their star,
where the ambient conditions could allow the presence of liquid water.
In particular, Earth-like landscapes with continents and shallow
water regions between water oceans appear to be a promising environment 
for life to form and evolve.
With the aim of finding potentially habitable exoplanets and indeed 
traces of extraterrestrial life, scientists are
(thinking about) designing and optimizing the next generation of space and 
ground-based telescopes and instruments
that would be able to characterize small, 
rocky exoplanets in the habitable zones of their stars 
\citep[see e.g.\ the recently published `Pathways to Discovery in Astronomy and Astrophysics for the 2020s' by the][]{NAP26141}.

The techniques that are currently the most successful 
in detecting such small, rocky exoplanets, i.e.\ the transit method
and the radial velocity method, give little
information about the planet's characteristics besides the (minimum) 
mass and the orbital parameters. The latter allow computing the
amount of incoming stellar flux and estimating, albeit very roughly,
the average surface temperature and the possible presence of liquid 
water on the planet and hence the possible presence of liquid 
surface water, if the planet is of the terrestrial type. 

A future technique to characterize the physical properties of
small planets is the direct detection of light that these planets
reflect as they orbit their parent star.  
This is very difficult because these planets are extremely faint 
compared to their parent star; 
the star will usually be at least $10$ million times brighter than 
the planet \citep[see][and references therein]{sphere,2021ExA...tmp..124S}.
The most promising candidates for direct observations of Earth-like
exoplanets orbiting solar-type stars are space telescopes, 
in particular when flown in combination with a star-shade. 
This architecture was originally proposed by \citet{star_shade_cash} 
and makes use of a star-shade spacecraft that blocks out the star's 
bright light, allowing a distant, formation flying space telescope 
to directly observe the reflected starlight of the orbiting planets. 
Two star-shade missions have been selected for study: the Star-shade
Rendez Vous Probe that aims to add a star-shade to 
the Nancy Grace Roman Space Telescope \citep{roman_starshade} 
(to be launched before 2030), 
and the HabEx mission 
\citep{hab_ex} that consists of a space telescope and a dedicated
star-shade (to be launched in the mid 2030s).
The next generation of ground-based telescopes, such as the 
Extremely Large Telescope 
(ELT)\footnote{\url{https://www.eso.org/public/teles-instr/elt/}} 
will also have the capability to directly image Earth-like exoplanets 
around nearby M-type stars but the contrast needed to detect signals of
such planets around Sun-like stars appears to be hard to achieve.

Apart from measuring the total flux of light that is reflected by
an exoplanet, some telescope designs also aim to measure the 
degree and direction of linear polarization of this light,
for example, the Nancy Grace Roman Space Telescope 
\citep{2021AAS...23732703M}, and it is being considered for HabEx and
LUVOIR (a UV-space telescope concept)
\citep[see][and references therein]{2021ExA...tmp..124S}.
Polarimetry for the characterization of small exoplanets,
in particular the detection of oceans, was explicitly mentioned in 
`Pathways to Discovery in Astronomy and Astrophysics for the 2020s' 
\citep[][]{NAP26141}.

Dedicated space and/or ground-based telescopes, with or without
polarimetric capabilities, are expected to 
take the first images of Earth-like exoplanets sometime in the next 
decades. However, due to the huge distances, these exoplanets will 
appear as single pixels in any such image, much like the Earth 
itself in the iconic Pale Blue Dot picture taken from about 
$6$ billion kilometers distance by Voyager 1.

\begin{figure}[t!]
\centering
\includegraphics[width = 87mm]{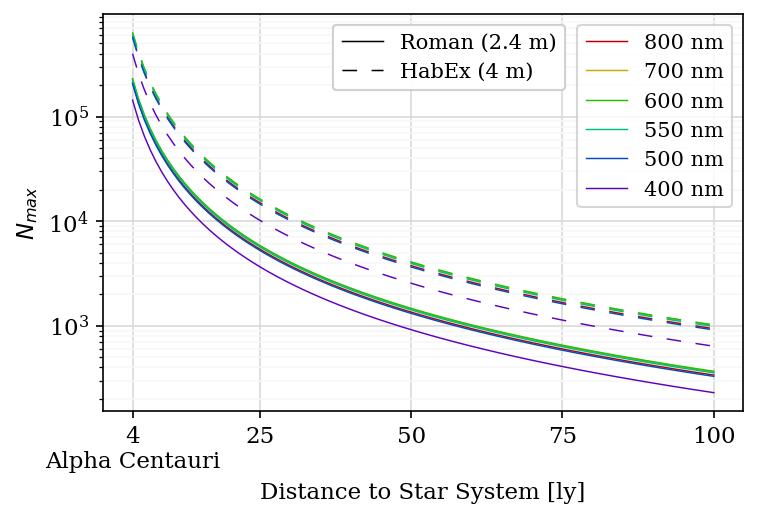}
\caption{The number of photons that would be detected by the 
        Nancy-Roman and HabEX space telescopes (apertures 
        $2.4~\text{m}$ and $4.0~\text{m}$, respectively) from a 
        white, Lambertian reflecting, Earth-sized planet in a 
        $1~\text{au}$ orbit around a solar-type star at $\alpha=0^\circ$,
        in optical spectral bands with widths of $50~\text{nm}$ 
        for a $3.0$~hour observation.
        These lines can be regarded as upper bounds for the 
        number of photons from an Earth-twin (the lines for
        the $500$, $550$, $600$, $700$, and 
        $800~\text{nm}$ bands are
        very close together).
        The SNR is proportional to $\sqrt{N_\text{max}}$.
        }
\label{fig_number_of_photons}
\end{figure}

Even with the largest telescope concepts, the photons received 
from a small exoplanet will have to be collected for up to hours
to make sure that the planet signal rises above the noise. 
In Fig.~\ref{fig_number_of_photons}, we show the number of 
photons that would be received by the Star-shade Rendez Vous 
Probe (with the Nancy Roman Space Telescope) and HabEx, 
which have apertures of $2.4$ and $4$ meters, respectively, 
for a completely white, Earth-sized planet around a solar-type 
star at full phase (the planet is then actually behind the star,
so this is a theoretical maximum value). 
We furthermore assumed that the total flux in a $50~\text{nm}$-wide 
spectral band is integrated over three hours.

%--------------------------------------------------------------------------
%--------------------------------------------------------------------------

%--------------------------------------------------------------------------
\begin{figure}[ht!]
\centering
\includegraphics[width= 0.9\linewidth]{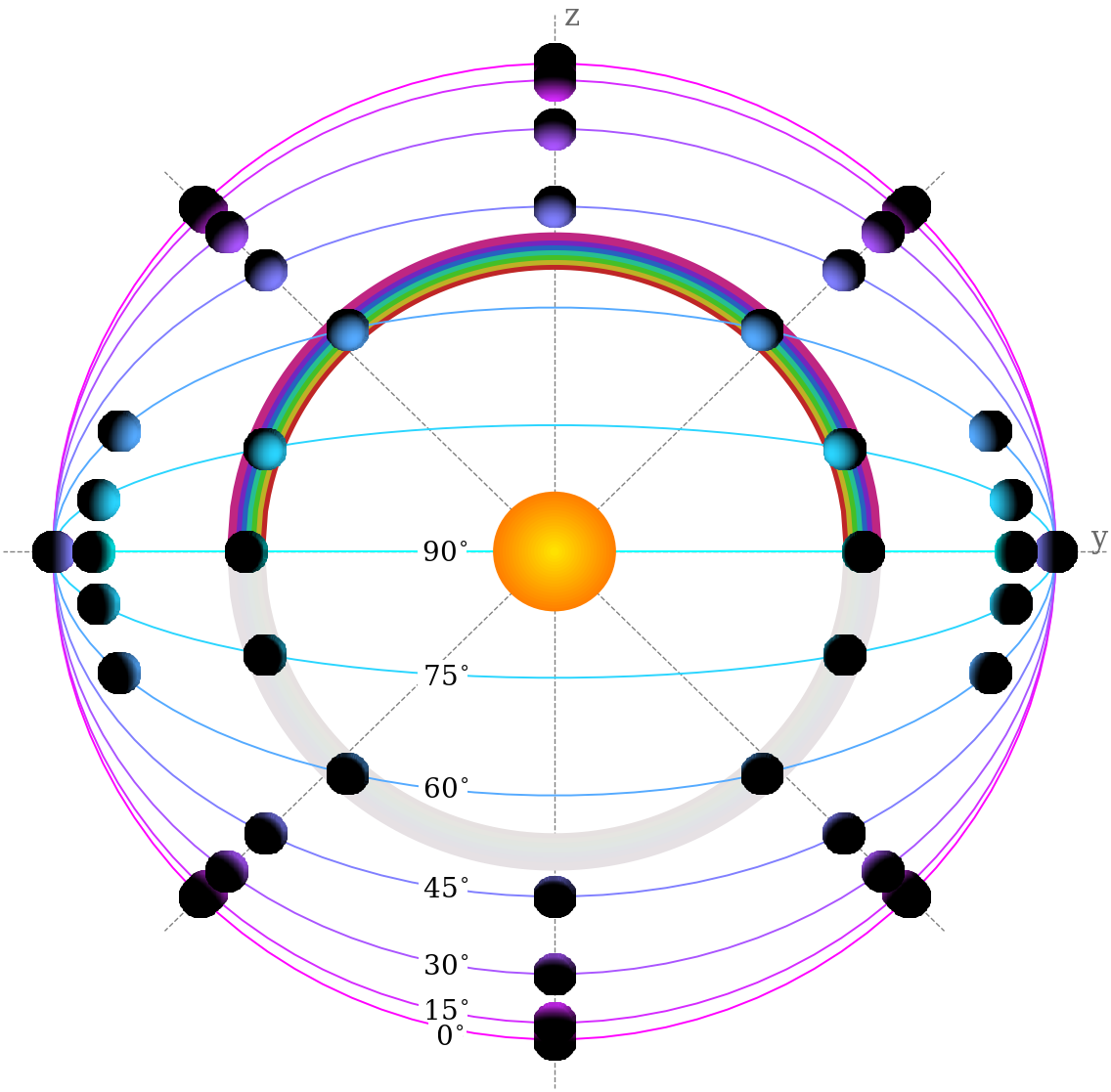}
\caption{Planetary orbits with inclination angles $i$ ranging from 
        0$^\circ$ to 90$^\circ$,
        with the lower half of the figure closest to the observer 
        (we thus define 
        $i$ as the angle between the orbit normal and the vector to the observer).
        Each planetary system is defined with respect to a Cartesian
        coordinate system $(x,y,z)$, with the $x$-axis pointing towards the 
        observer, and the $y$- and $z$-axis as shown.
        Points where the planets have the same phase angle, $\alpha$ 
        (defined as the angle between the vectors from the planet towards 
        the star and the observer), form circles, because they lie on a 
        cone with a tip-angle equal to $\alpha$ and also 
        on a sphere with the radius of the (circular) orbits. The 
        rainbow is thus a semicircle, since it is the intersection at
        $\alpha=38^\circ$. The color sequence of the rainbow 
        is reversed compared to the rainbow that forms in rain 
        droplets \citep{2012A&A...548A..90K}.
        Planets are shown at the eight locations 
        where we assume observations would be taken.
        }
\label{fig_orbital_positions}
\end{figure}
%-------------------------------------------------------------------------

If the planet's surface is horizontally inhomogeneous
and/or if the planet has a broken cloud-deck, its total flux 
signal as well as its polarization signal will vary as the planet
rotates about its axis and different parts of its surface are 
illuminated by the light of its parent star.
The planet's signal will thus depend on the longitudinal 
variation over the planet.
Also, as the planet orbits the star, its phase will change, just
like the lunar phase changes throughout a month. 
With an exoplanet, we will, however, not see a change in the 
shape of the illuminated disk, as we do for the Moon, but a change 
in the overall total and polarized fluxes of the unresolved pixel.
As can be seen in Fig.~\ref{fig_orbital_positions},
the range of phase angles that the planet will cover along its orbit depends on the inclination angle $i$ of the orbit: if the orbit is seen
edge-on \textbf{($i=90^\circ$)}, the phase angle $\alpha$ varies between almost $0^\circ$ to $180^\circ$,
while if the orbit is face-on \textbf{($i=0^\circ$)}, 
the phase angle is always $90^\circ$.
From which part of the exoplanet an observer will receive reflected
starlight at any given phase angle also depends on the orientation
of the rotation axis of the planet.
If the axis is tilted with respect to the planet's orbital plane, 
the sub-stellar point travels between the tropical circles. 
This results in a reflected light signal that depends on the surface
and cloud variation with latitude.

Because the reflection properties of a planetary surface and 
atmospheric constituents will also depend on the wavelength, 
the reflected light signal of a rotating planet will thus depend
on time and also on the wavelength, 
\citep[][and references in the latter]{2001Natur.412..885F,colors_of_earth}.
Measurements of the temporal and spectral variations of the signal
of a single pixel exoplanet contain information about the horizontal
variability of the planet.
How to retrieve a surface map of an exoplanet from reflected 
total flux light curves has been studied by 
\citet{fujii_2012,sot_dynamic,2D_alien_map,exocartographer,nn_cartography}.
They have shown that it is indeed possible to constrain a planet's 
rotation axis and retrieve a rough planetary surface map. 
In these studies, the planet's surfaces are treated as Lambertian 
reflectors, which diffusely reflect starlight in all directions.

Lambertian reflectors do not include polarization.
As the total flux of the reflected starlight depends on the 
properties of the illuminated and visible part of the planetary disk,
so does the linear polarization signal of this light. 
While the starlight that is incident on the planet can be assumed 
to be unpolarized, 
Rayleigh scattering by the gases in the atmosphere, scattering by cloud
particles and reflection by surfaces like oceans will usually 
polarize the reflected light,
depending on the directions of the incident and reflected light,
and hence depending on the phase angle and the location on the planet
\citep[for polarization signals of Earth-like planets, see][and references therein]{stam_2008,colors_of_earth,blue_white_red}.

Here, we propose a retrieval algorithm that uses light curves computed 
assuming bidirectional reflection, as described by \citet{pymiedap} and \citet{blue_white_red}.
The radiative transfer computations fully include polarization 
and multiple scattering and are performed for an Earth-like atmosphere 
with Rayleigh scattering, water-clouds that create a rainbow at 
certain phase angles and thus orbital locations 
(Fig.~\ref{fig_orbital_positions} shows such locations along the 
planetary orbits) and oceans 
that exhibit the bright glint pattern due to specular reflection. 
The glint significantly affects a planet's bidirectional reflection
curve as compared to Lambertian reflection. 

The use of bidirectional and polarized light curves requires a new 
approach to planet mapping. 
Since bidirectional reflection is more complicated than Lambertian
reflection, analytical methods would be difficult to apply 
so different tools are required. We use neural networks, which have several 
advantages: they are universal approximators \citep{HORNIK1991251} 
and can thus find a planet's rotation axis and they have
achieved state-of-the-art-results in classification problems 
\citep{nn_overview}, and surface mapping is at heart a classification 
problem.
Our neural networks are designed to use 
the temporal variations of the reflected light signal of an exoplanet 
to retrieve planet characteristics such as the orientation of the 
rotation axis, the surface and the cloud coverage.
Our networks are trained with a large set of simulated observations 
of various model planets before being provided with the 
validation data set, i.e.\ simulated observations 
of model planets that were not in the training set.

In Sect.~\ref{section_numericalmethod}, we describe the characteristics
of our model planets including the surface types that we consider,
and we explain the numerical method that we use to 
simulate the total and polarized flux curves of our model planets,
and the model observations, including photon noise, 
in the training and the test data sets.
In Sect.~\ref{sect_neural_networks}, we describe the general 
architectures of our neural networks, how they compare to other
networks that have been applied in similar research questions,
and how we train and validate the network.
In Sects.~\ref{sect_results_rotation_axis}, \ref{sect_albedo_maps}, 
and~\ref{sect_surface_type_maps}, we describe the specific 
neural network architectures and the results for the retrievals of 
planetary rotation axes, albedo maps, and surface type maps, 
respectively. 
In these sections, we also investigate the retrieval errors
that result from training a neural network on Lambertian reflecting 
planets and applying it then on realistically bidirectionally
reflecting test planets, and we discuss the benefits of including
polarization.
In Sect.~\ref{sect_conclusions}, we provide our conclusions
and a number of recommendations for future work.

%%%%%%%%%%%%%%%%%%%%%%%%%%%%%%%%%%%%%%%%%%%%%%%%%%%%%%%%%%%%%%%%%%%%%%%%%%%%%%
\section{Method: radiative transfer}
\label{section_numericalmethod}

%%%%%%%%%%%%%%%%%%%%%%%%%%%%%%%%%%%%%%%%%%%%%%%%%%%%%%%%%%%%%%%%%%%%%%%%%
\subsection{Total and polarized fluxes}
\label{sect_fluxes_phase_curves}

We provide our neural network with simulated observations of the total 
and polarized fluxes of starlight reflected by an exoplanet.  
These fluxes form a so-called Stokes vector 
\citep[see][]{scattering_in_atmospheres}:
\begin{equation}
    \vec{I} =
    \begin{bmatrix} I \\ Q \\ U \\ V \end{bmatrix} ,
\label{eq_stokesvector}
\end{equation}
with $I$ the total flux, $Q$ and $U$ the linearly polarized 
fluxes, $(Q^2 + U^2)^{1/2}$
the total linearly polarized flux, 
and $V$ the circularly polarized flux.

The reflected vector $\vec{I}$ that 
arrives at a distant observer pertains to the planet as a single 
point of light, and thus comprises the locally reflected light
integrated over
the illuminated and visible part of 
the planetary disk. 
Given a wavelength $\lambda$ and a planetary phase angle $\alpha$,
vector $\vec{I}$ can be computed using
\citep[see Eq.~(5) in][]{2004A&A...428..663S}
\begin{equation}
   \vec{I}(\lambda,\alpha) = 
           A_{\rm G}(\lambda) \hspace*{0.1cm}
           \vec{R}(\lambda,\alpha) 
           \hspace*{0.1cm} \frac{r^2}{D^2}
           \hspace*{0.1cm}
           \pi F_0(\lambda) .
\label{eq_reflectedstokesvector}
\end{equation}
Here, $A_{\rm G}$ is the planet's geometric albedo, which generally
depends on $\lambda$ through the spectral dependence of the
planet's reflective properties (for a horizontally inhomogeneous
planet, this will depend on which part of the planet is turned towards
the observer), $r$ is the radius of the planet, 
and $D$ is the distance between the planet and the observer. 
Furthermore, %and
$\pi F_0$ is the stellar flux %that is
incident on the planet, which we assume to be unpolarized and unidirectional.
Furthermore, $\vec{R}$ is the vector describing the angular variation of the starlight %that is
reflected by the planet
(this is represented by a vector instead of a matrix because we assume the 
%incoming
starlight to be unpolarized); $\vec{R}$ 
depends on the characteristics of the planetary surface and atmosphere, 
the wavelength $\lambda$, and the illumination and viewing geometries.

Because the circularly polarized flux $V$ of Earth-like planets is
approximately $5$~orders of magnitude smaller than the total flux $I$
\citep[see e.g.][]{colors_of_earth}, 
its detection is virtually impossible.
We therefore ignore $V$ in our simulations.
The other vector elements are normalized such that $I$ equals 
one at a phase angle of $0^\circ$, and we thus use
$A_{\rm G} I$ as the planet's (flux) phase curve. 
Next, we describe the surface and atmosphere characteristics of 
our model planets.

%-------------------------------------------------------------------------
% Figure 3:
%-------------------------------------------------------------------------
\begin{figure}[b!]
\centering
\includegraphics[width=87mm]{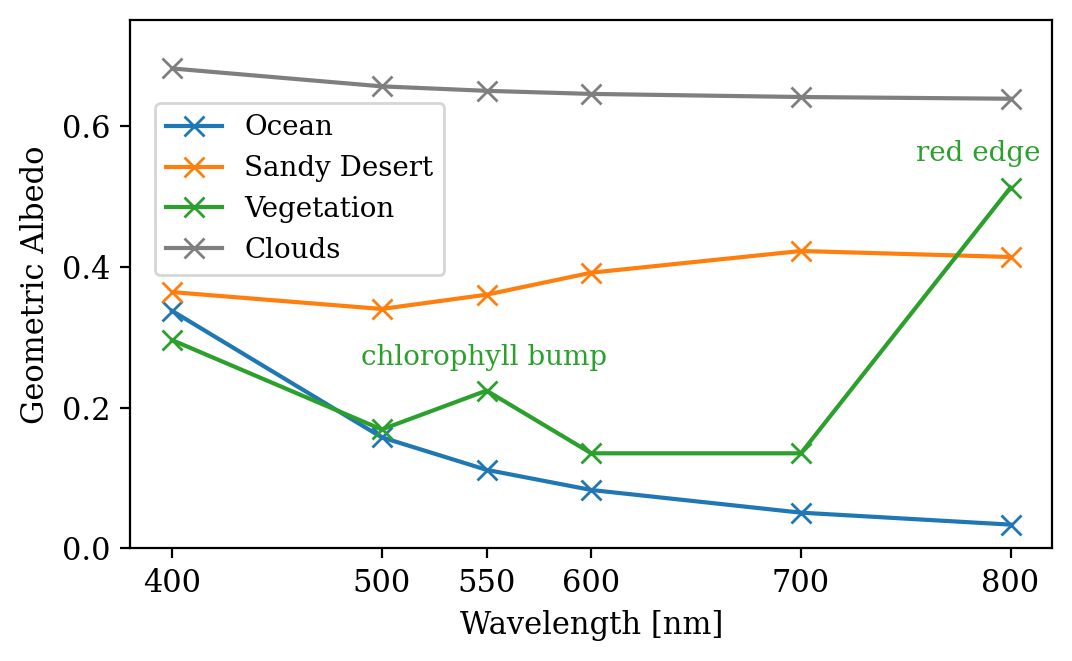}
\caption{The geometric albedos $A_{\rm G}$ of model planets
        that are completely covered by ocean (blue), sandy desert (orange),
        vegetation (green), or clouds (grey). Rayleigh scattering by the 
        gaseous atmosphere is included.
        }
\label{fig_surface_albedos}
\end{figure}
%-------------------------------------------------------------------------

%-------------------------------------------------------------------------
% Figure 4:
%-------------------------------------------------------------------------
\begin{figure*}[t!]
\includegraphics[width=160mm]{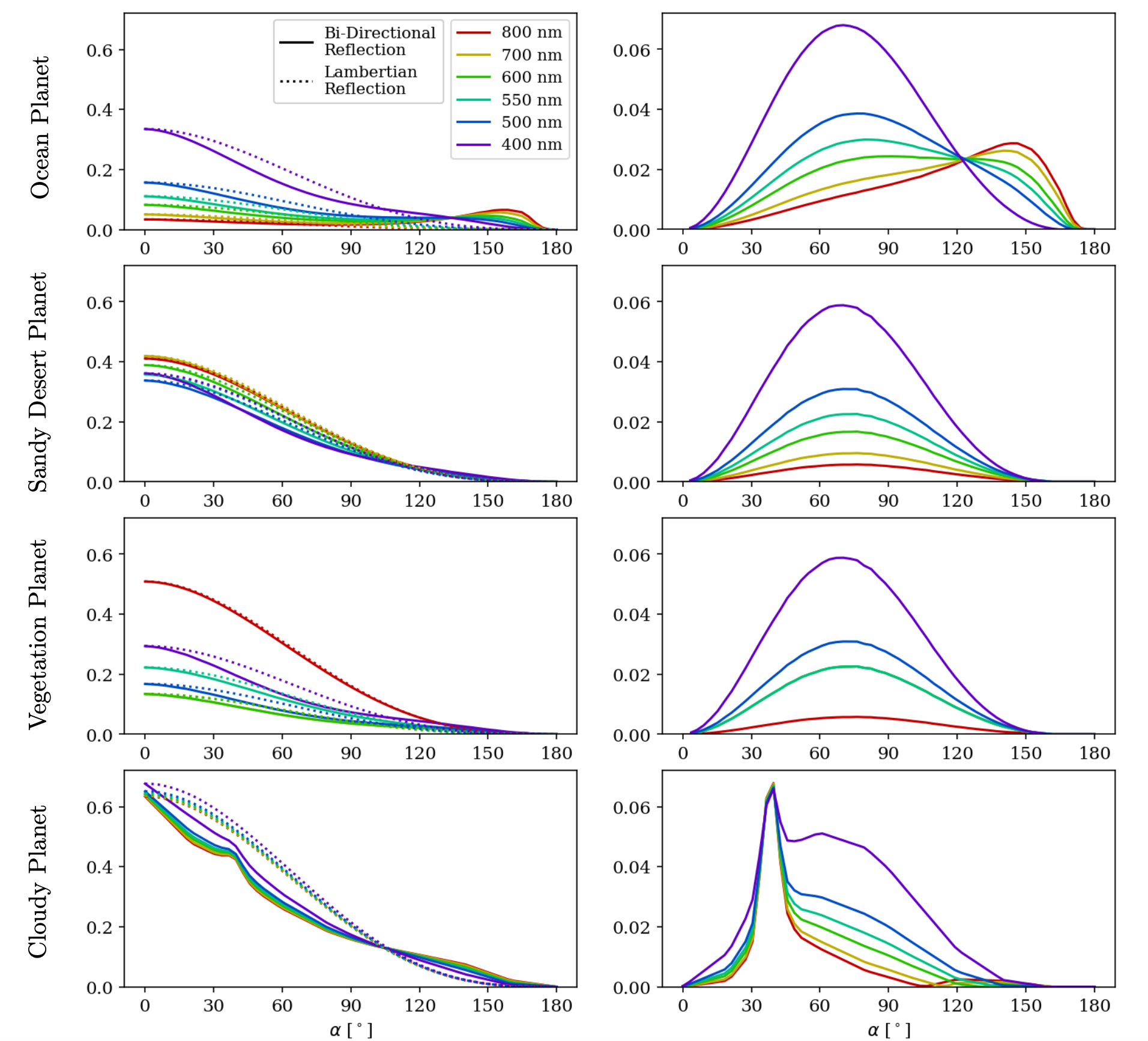}
\caption{Total flux (left column) and polarized flux (right column) 
         phase functions of horizontally homogeneous model planets 
         covered by ocean (first row), sandy desert (second row),
         vegetation (third row), or clouds (fourth row), 
         from $400$ to $800~\text{nm}$.
         The total fluxes have been computed for Lambertian
         reflecting (dotted lines) and bidirectionally
         reflecting planets (solid lines), and the polarized fluxes
         only for bidirectionally reflecting planets. 
         The curves are normalized such that a white Lambertian 
         reflecting planet at $\alpha=0^\circ$ has a total flux of one.
         Near $\alpha=180^\circ$, the approximation that the surface 
         and atmosphere are locally plane-parallel breaks down
         (because of the very low fluxes, this is not visible
         in the curves).
         }
\label{fig_phase_curves}
\end{figure*}
%-------------------------------------------------------------------------

%-------------------------------------------------------------------------
% Figure 5:
%-------------------------------------------------------------------------
%\input{fig_resolved_fluxes}
\begin{figure}[t!]
\includegraphics[width=87mm]{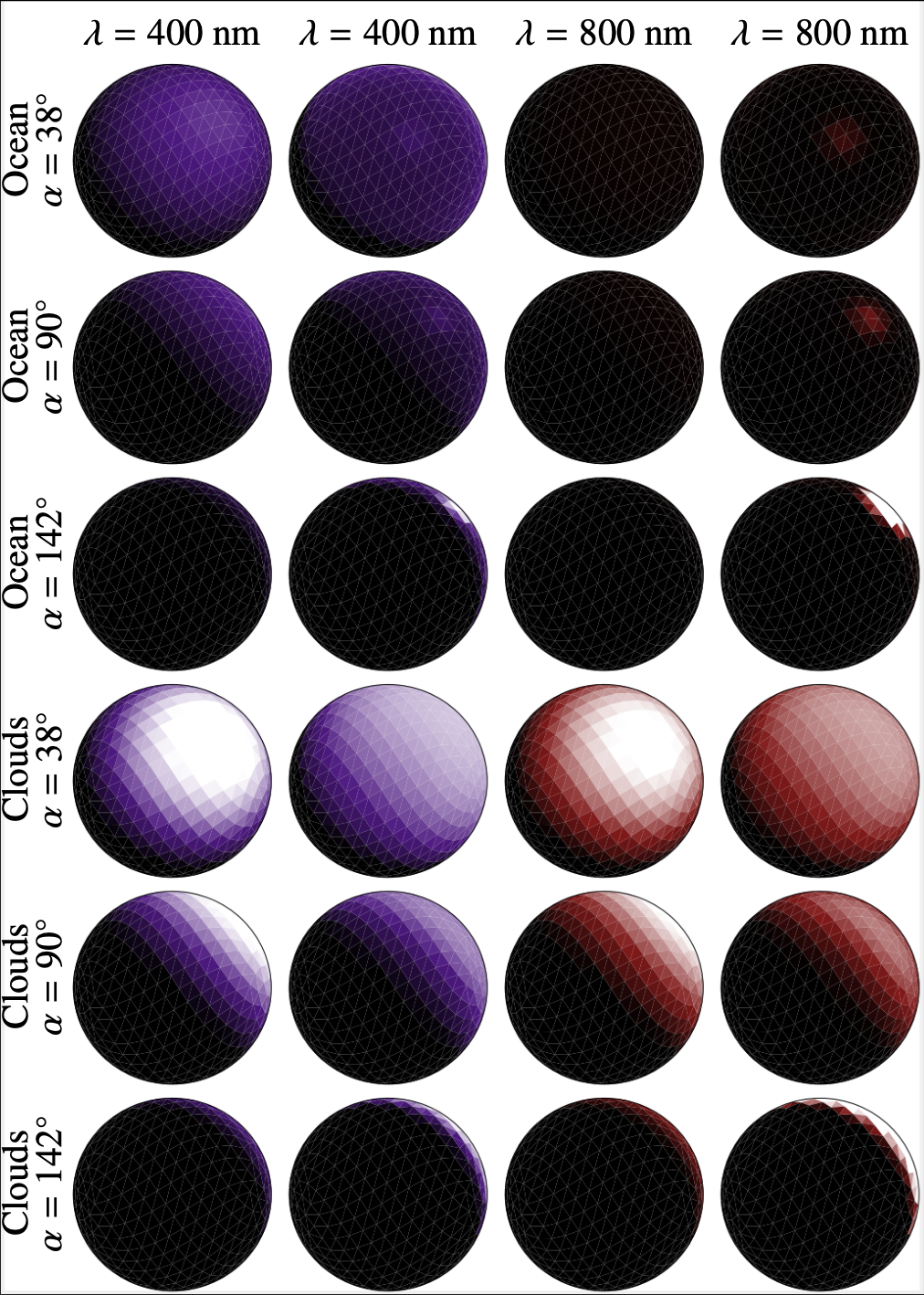}
\caption{Images of homogeneous, Lambertian and 
        bidirectionally reflecting planets for $\lambda=400$
        and $800~\text{nm}$, with the star to the top-right of each planet.
        A white facet corresponds to a brightness of $\geq 0.62$ 
        (normalised to a white, Lambertian reflecting facet).
        The overall brightest facet (i.e.\ $2.7$) is in the ocean 
        glint for $\alpha = 142^\circ$ and $\lambda = 800~\text{nm}$. 
        At the same $\alpha$ and $\lambda$, the cloudy facets 
        reach a maximum brightness of $1.1$.}
\label{fig_resolved_fluxes}
\end{figure}
%-------------------------------------------------------------------------

%%%%%%%%%%%%%%%%%%%%%%%%%%%%%%%%%%%%%%%%%%%%%%%%%%%%%%%%%%%%%%%%%%%%%%%%%%%%
\subsection{Phase curves for homogeneous planets}
\label{sect_surfaces_atmospheres}

Like Earth, our model planets are covered by
different surface types and have gaseous atmospheres that
can contain liquid water clouds. 
Our radiative-transfer algorithm \citep{pymiedap}, 
computes the starlight that is reflected by a locally plane-parallel 
and horizontally homogeneous surface-atmosphere model,
where the atmosphere consists of a stack of homogeneous layers.
To capture the spatial variation in surface-atmosphere models
while adhering to the requirements of our radiative transfer algorithm,
we describe each spherical planet with locally flat facets  
each of which is assigned a specific surface-atmosphere model. 

We use three surface types: ocean, sandy desert, and vegetation.
We do not include ice for polar caps and we limit ourselves to a static model. 
Although ice caps can be permanent, a realistic model would account for seasonal
growing and melting depending on the 
obliquity of the planet's rotation axis. Studying the retrieval of
such seasonal effects is beyond the aims of this work
(see the recommendations in Sect.~\ref{sect_conclusions}).

The reflection by a given facet depends not only on the local surface-atmosphere model and wavelength $\lambda$, but also on
the local zenith angle of the star, the local zenith angle of the
observer, and the azimuthal angle between the directions towards
the star and the observer. These angles depend on the facet's location 
on the planet and on the phase angle $\alpha$.

In the papers about exoplanet cartography by
\citet{fujii_2012,sot_dynamic,2D_alien_map,exocartographer,nn_cartography},
the model planets reflect Lambertian, i.e.\ isotropic
and unpolarized.
To allow for a comparison with results in those papers, we implemented  %such
Lambertian reflecting model planets as well.
However, we also implemented bidirectional and polarized reflection 
using an efficient adding-doubling radiative
transfer algorithm and the summation of local reflection vectors 
across the planetary disk as described by 
\citep[][]{pymiedap,blue_white_red,colors_of_earth}.\footnote{In these papers, the facets that describe the variation across the planet are determined by a grid of square pixels modeling a distant detector, while our Fibonacci sphere facets are defined on the planet surface.}
This allows for more accurate simulations of the reflected fluxes angular reflection patterns due to the clouds, such as 
rainbows, the ocean glint, and, indeed, polarization.

Our Earth-like model atmosphere consists of 
layers that contain an Earth-like gas-mixture and, optionally, 
cloud particles. For the properties of the anisotropic Rayleigh
scattering gas and the clouds that consist of 
Mie-scattering liquid water particles, see \citet{stam_2008}. 
We use the reflection by the rough ocean surface as implemented by 
\citet{blue_white_red}, assuming a wind-speed of $7~\text{m/s}$.
The sandy desert and vegetation surfaces are modelled as 
Lambertian reflectors with wavelength dependent 
surface albedos \citep[see][]{colors_of_earth}.

Figure~\ref{fig_surface_albedos} shows the geometric albedos
$A_\text{G}$ (the phase functions $A_\text{G}I$ 
at $\alpha=0^\circ$) of 
four model planets that are completely covered by ocean, sandy
desert, vegetation, or clouds, all computed at 
$\lambda=400$, $500$, $550$, $600$, $700$, and $800~\text{nm}$. 
At $400~\text{nm}$, the relatively high 
values for $A_\text{G}$ of the cloud-free planets are due to 
Rayleigh scattering by the atmospheric gas. With increasing 
$\lambda$, the gas optical thickness decreases allowing the surface
albedo to dominate the signal. Because the ocean is very dark,
the albedo of the ocean-covered planet decreases with increasing $\lambda$.
At $550~\text{nm}$, the chlorophyll signature of the vegetation is 
captured, 
% \textbf{Of course this is typical for plants on Earth.}
and the strong increase of $A_\text{G}$ of the vegetation planet 
above $700~\text{nm}$ is due to the so-called red edge that is characteristics 
of terrestrial vegetation \citep[see e.g.][]{vegetation_biosignatures}. 
The highly reflective clouds dominate the signal of the cloudy
planet at all wavelengths.

Figure~\ref{fig_phase_curves} shows the total and polarized fluxes
of the horizontally 
homogeneous planets with ocean, desert, vegetation, and clouds 
from Fig.~\ref{fig_surface_albedos}, as functions of the phase angle
$\alpha$.
The curves are shown both for the Lambertian reflecting model
planets (thus without polarized fluxes) and for the bidirectionally
reflecting planets, for wavelengths from $400$ to $800~\text{nm}$.
For ease of comparison between the directional and Lambertian reflection
models, the geometric albedos of each Lambertian reflecting model planet are normalized to the 
corresponding bidirectionally reflecting planet.

The Lambertian planets are all brighter than
their corresponding bidirectionally reflecting planets between about
$\alpha=10^\circ$ and $100^\circ$.
At the largest phase angles, they tend to be darker, in particular 
the ocean covered planet. Indeed, due to the ocean glint, the
bidirectionally reflecting model planet is relatively bright around
$\alpha=160^\circ$, especially at the longest wavelengths
\citep[for a more in-depth discussion of the reflected light signals of ocean planets, see][]{blue_white_red}.

The bump at $\alpha=38^\circ$ in the total flux of 
the (bidirectionally reflecting) cloudy planet is the rainbow, 
which is characteristic for liquid water clouds.
We do not have a bidirectional reflection model for
the desert and vegetation surfaces. Therefore, 
the curves for those model planets virtually coincide with those of 
the Lambertian planets at the longer wavelengths. This is because
at those wavelengths, 
the atmospheric Rayleigh scattering optical thickness is 
very small and the surface reflection dominates the signal.

Figure~\ref{fig_resolved_fluxes} illustrates the differences 
between Lambertian and bidirectional reflection on spatially 
resolved ocean and cloud-covered planets at different phase
angles and at wavelengths of $400$ and $800~\text{nm}$.
On the ocean planet, the glint pattern is strongest at $\alpha=142^\circ$ and 
$\lambda= 800~\text{nm}$, as expected from Fig.~\ref{fig_phase_curves}.
At $400~\text{nm}$, the glint pattern is weakened  
due to Rayleigh scattering. At smaller phase angles, the Rayleigh
scattering spreads out the reflected flux across the planetary 
disk when compared to the Lambertian reflection. 
A similar evening-out of the reflected fluxes can be seen for the
cloudy model planet: the Lambertian reflecting model planet is much
brighter across the sub-solar region than the bidirectional 
reflecting planet. 
At the largest phase angles, the forward scattering behaviour of the
cloud particles leaves the bidirectional reflecting planet much 
brighter than the Lambertian reflecting planet. 

These differences in the brightness distribution across the planetary
disk will influence the retrieval of the map of the planet:
an algorithm that assumes a Lambertian reflecting model
planet while the actual planet reflects bidirectional is thus
expected to retrieve lower surface albedo's in the sub-solar region.
Consequences of assumptions on the retrievals by our
neural network will be discussed further in this paper.

%-------------------------------------------------------------------------
% Figure 6:
%-------------------------------------------------------------------------
\begin{figure}[t!]
\centering
\includegraphics[width=87mm]{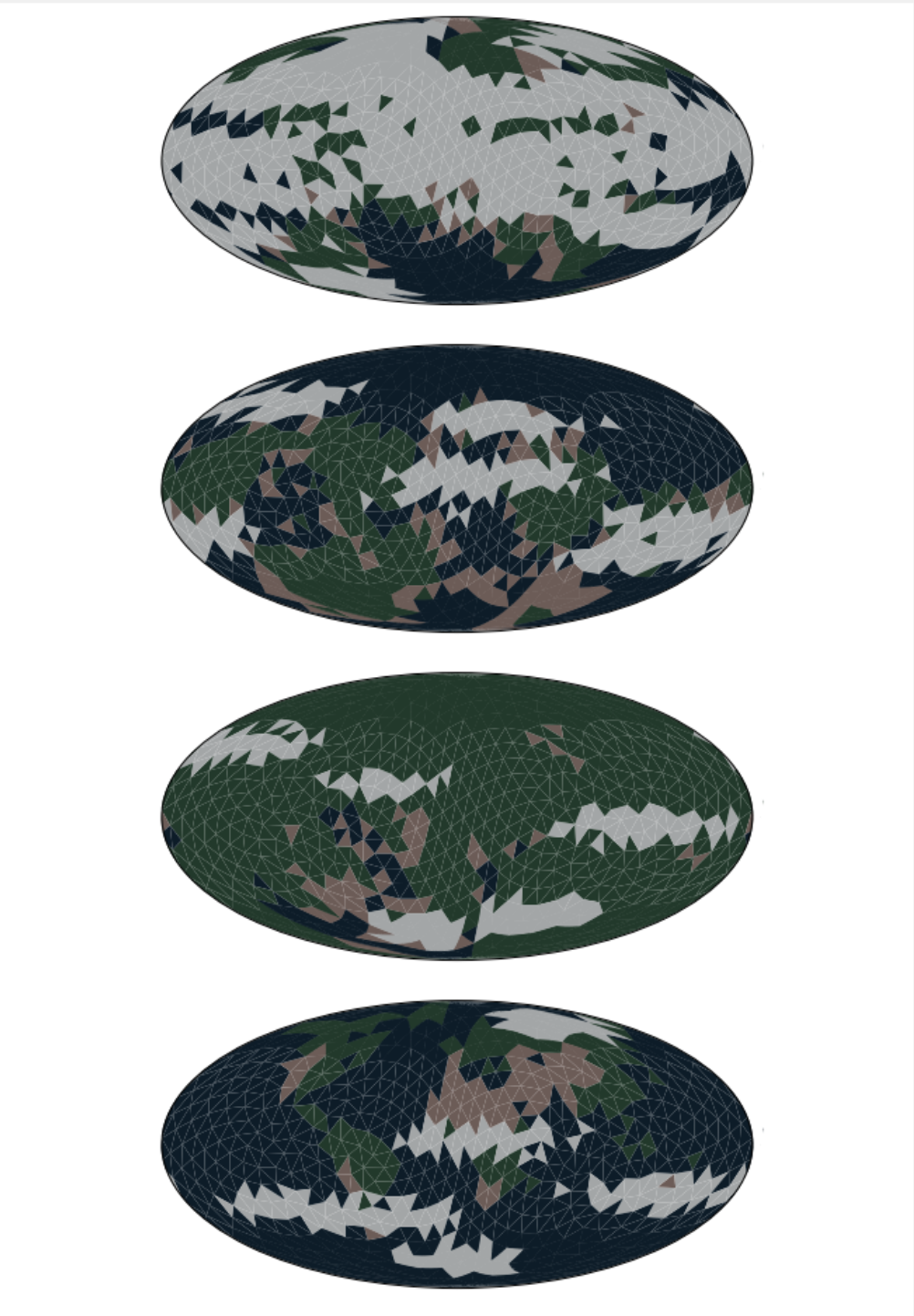}
\caption{Examples of model-planet maps used to train the neural network, 
         including a model Earth (bottom) 
         with $(x,y,z)\approx (71\%,13\%,16\%)$ in Eq.~\ref{eq:abc}, 
         and a selected cloud pattern. The flux curves of the model Earth
         in an edge-on orbit ($i=90^\circ$) are shown in 
         Fig.~\ref{fig_earth_curves}. 
         The RGB color values of each surface type are given 
         by the albedos at the wavelengths of 700, 550, and 500 nm 
         (see Fig.~\ref{fig_surface_albedos}), respectively.}
\label{fig_earth_with_clouds}
\end{figure}
%-------------------------------------------------------------------------

%%%%%%%%%%%%%%%%%%%%%%%%%%%%%%%%%%%%%%%%%%%%%%%%%%%%%%%%%%%%%%%%%%%%%%%%%%
\subsection{Surface-type and cloud maps}

The positions of the facets on the planet are determined by 
the Fibonacci sphere \citep{fibonacci} instead of the 
HEALPix scheme \citep{gorski_2005}.
The Fibonacci sphere has the advantage that it allows any (even) number 
of approximately equal-area and equilateral facets. 
Through testing, we determined 
that 1000 facets covering the whole planet strikes a good balance 
between capturing surface and cloud patterns while ensuring the
efficiency of training the network.

To create model planets with various coverage fractions of different
surface types, we use the following probability distribution for 
fraction $x$ (ocean), $y$ (desert), and $z$ (vegetation):
\begin{equation}
   p(x,y,z) = \frac{\delta(1-x-y-z)}{2\pi \!\sqrt{xyz}}.
\label{eq:pxyz}
\end{equation}
Here, $\delta$ is the Dirac-delta function. The unconditional distributions for having a fraction $x$ ocean, 
and for having respective fractions 
$x$ and $y$ for ocean and desert are then
\begin{equation}
p(x)= \frac{1}{2 \!\sqrt{x}} , \hspace*{0.1cm} p(x,y)=
\left\{
\begin{array}{ll}
\dfrac{1}{2\pi \!\sqrt{xy(1-x-y)}} & \text{if} \hspace*{0.1cm} x + y < 1 \\
0 & \text{else.}
\end{array} \right.
\label{eq:pxy}
\end{equation}
The prior probability distribution in Eq.\ (\ref{eq:pxyz}) 
is symmetric in the three variables and, due to it's singular 
behavior at the edges, will generate
a significant number of planets with
one dominant surface type. We thus train the network to also recognize 
planets that are mostly covered with one of the surface types, 
like water worlds or desert planets.

To create a surface map on the sphere,
we draw numbers $q_1$, $q_2$ from the uniform 
distribution on $[0,1]$ and compute the fractions as follows:
\begin{equation}
    x = q_1^2 , \quad
    y = \tfrac{1}{2} \hspace*{0.05cm} (1 - x) \hspace*{0.05cm} (1 - \cos\pi q_2) , \quad
    z = 1 -  x - y.
\label{eq:abc}
\end{equation}
We then assign to each facet a fictional elevation using the tetrahedral 
subdivision method (with an offset exponent $q$ of 0.5) that was 
originally designed for video game applications \citep{map_generation}. 
We create surface maps by establishing the elevation cut-off levels 
of each surface type based on the coverage fractions $x$, $y$, and 
$z$ with the ocean at the lowest elevations, vegetation at the highest, 
and desert at those in between.

The clouds overlay the surface and to select the facets of our 
model planets that are cloud covered, we draw a cloud 
coverage fraction from the probability distribution in Eq.~(\ref{eq:abc}),
and then use the method described in \citet{pymiedap} to create
patchy, zonal cloud patterns. 
The clouds have an optical thickness of $10$ at $550~\text{nm}$,
and the surface below them is assumed to be black.
The clouds are static with respect to the surface across all epochs. %{\color{red}{say something about time variations of the clouds}}
A sample model planet with a surface coverage resembling 
Earth, thus with $(x,y,z)\approx (71\%,13\%,16\%)$, is shown in 
Fig.~\ref{fig_earth_with_clouds}.

Once we have the planet's surface and cloud map, we compute the
starlight that is reflected by each facet and, by summing the 
contributions of the individual facets, by the planet as a whole
\citep[see][]{pymiedap}.

%-------------------------------------------------------------------------
% Figure 7:
%-------------------------------------------------------------------------
\begin{figure}[htb!]
\centering
\includegraphics[width=100mm]{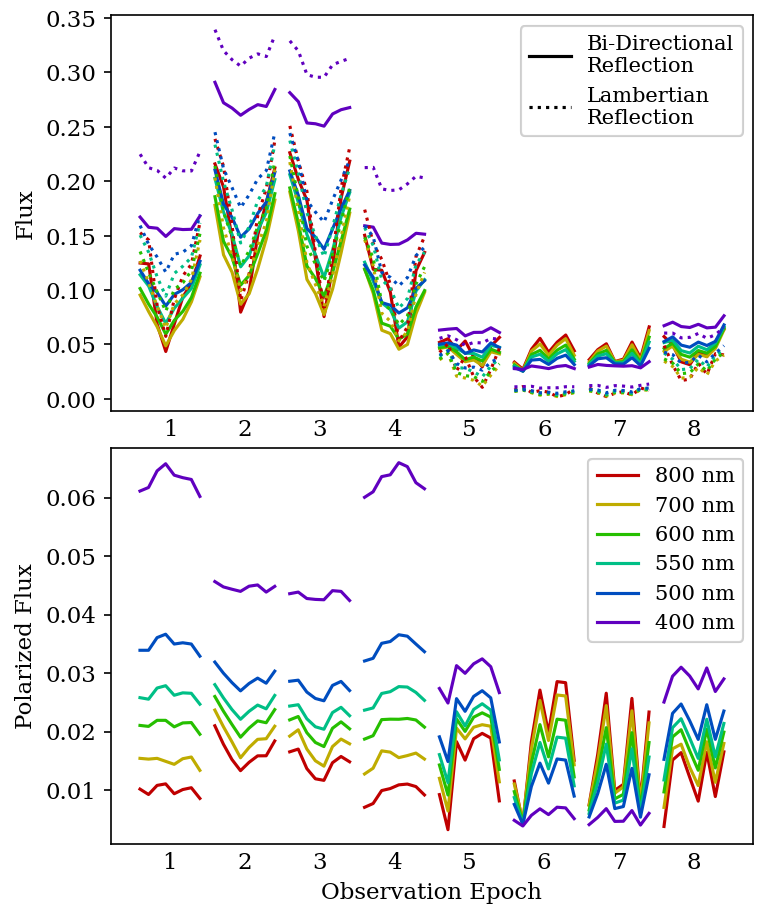}
\caption{Total (top) and polarized (bottom) flux curves for our model Earth
         (bottom of Fig.~\ref{fig_earth_with_clouds}) in an 
         edge-on orbit with 
         an axial tilt angle of $90^\circ$ 
         (see Table~\ref{tab_orbital_positions} for the planet's position 
         and phase angle during each observational epoch).
         The polarized flux is calculated only for the bidirectionally 
         reflecting planets, since Lambertian reflection is unpolarized.}
\label{fig_earth_curves}
\end{figure}
%-------------------------------------------------------------------------
%-------------------------------------------------------------------------
% Table 1, with angles and different observational epochs:
%-------------------------------------------------------------------------
\begin{table}[b]
\caption{Phase angle $\alpha$ vs. observational epoch and orbital 
         inclination.}
\begin{tabular}{c|cccccccc}
\hline\hline
\setlength\tabcolsep{0.2pt}
$i$ & 1 & 2 & 3 & 4 & 5 & 6 & 7 & 8 \\
\hline
 $\phantom{0}0^\circ$ &  $90^\circ$ & $90^\circ$ & $90^\circ$  & $90^\circ$ & $\phantom{0}90^\circ$ & $\phantom{0}90^\circ$ & $\phantom{0}90^\circ$ & $\phantom{0}90^\circ$ \\
 $15^\circ$ & $90^\circ$ & $79^\circ$ & $75^\circ$  & $79^\circ$ & $\phantom{0}90^\circ$ & $101^\circ$ & $105^\circ$ & 101$^\circ$ \\
 $30^\circ$ & $90^\circ$ & $68^\circ$ & $60^\circ$ & $68^\circ$ & $\phantom{0}90^\circ$ & $112^\circ$ & $120^\circ$ & $112^\circ$ \\
 $45^\circ$ & $90^\circ$ & 55$^\circ$ & $45^\circ$ & $55^\circ$ & $\phantom{0}90^\circ$ & $125^\circ$ & $135^\circ$ & $125^\circ$ \\
 $60^\circ$ & $65^\circ$ & \includegraphics[height = 2.6mm]{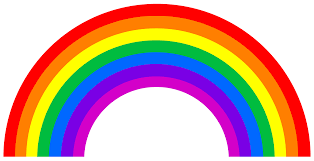} & \includegraphics[height = 2.6mm]{rainbow_symbol.png} & $65^\circ$ & $115^\circ$ & $142^\circ$ & $142^\circ$ & $115^\circ$ \\
 $75^\circ$ & $67^\circ$ & \includegraphics[height = 2.6mm]{rainbow_symbol.png} & \includegraphics[height = 2.6mm]{rainbow_symbol.png}  & $67^\circ$ & $113^\circ$ & $142^\circ$ & $142^\circ$ & $113^\circ$ \\
 $90^\circ$ & $67^\circ$  &  \includegraphics[height = 2.6mm]{rainbow_symbol.png} & \includegraphics[height = 2.6mm]{rainbow_symbol.png} &  $67^\circ$ & $113^\circ$ & $142^\circ$ & $142^\circ$ & $113^\circ$ \\
\hline
\end{tabular}
\tablefoot{The \includegraphics[height = 2.6mm]{rainbow_symbol.png}
symbol indicates that
$\alpha=38^\circ$ and that the rainbow feature should be visible (provided
the planet has liquid water clouds). See Fig.~\ref{fig_orbital_positions} 
for the geometries at the given orbital inclination angles $i$. 
}
\label{tab_orbital_positions}
\end{table}
%-------------------------------------------------------------------------

%%%%%%%%%%%%%%%%%%%%%%%%%%%%%%%%%%%%%%%%%%%%%%%%%%%%%%%%%%%%%%%%%%%%%%%%%%
\subsection{Orbits and rotation axes of the model planets}
\label{sec_orbital_parameters}

Exoplanets move in planes that are generally observed at an angle. 
The inclination angle $i$ of this orbital plane is defined as the 
angle between the normal on the plane and the direction towards 
the observer. 
As the planet orbits its star, $\alpha$ ranges from 
$(90^\circ-i)$ to $(90^\circ+i)$.
Figure~\ref{fig_orbital_positions} illustrates planetary 
orbits for the values of $i$ that we use to simulate the 
observations of the model planets for training and testing the network:
$i$ ranges from $0^\circ$ to $90^\circ$ with 
steps of $15^\circ$. Since we consider 
directly detected exoplanets, the inclination
angles of their orbits could be obtained from the observations 
and will be provided to the network.

Rather than continuous measurements throughout a complete orbit, 
we assume that each model planet is observed at eight 
locations along its orbit, see Fig.~\ref{fig_orbital_positions}.
For our application, we assume that an exoplanet 
will be observed spatially resolved from the stellar flux,
as the latter will be several orders of magnitude larger and exhibit
its own temporal variations. 
We thus exclude small and large phase angles, where the angular
distance between the planet and the star is smallest, from our
simulated observations, and only include
locations where $38^\circ \leq \alpha \leq 142^\circ$.
For real observations, the limiting phase angle range, 
or the inner working angle (IWA), depends on the
telescope and for example on the use of a coronagraph or a star shade, 
and on the characteristics of the planetary system such as the
angular distance between the planet and the star.
With our choice, we are on the safe side.

The lower limit of $38^\circ$ ensures the visibility
of the rainbow of starlight that has been scattered by liquid 
water cloud particles, which could be used for the characterization of 
clouds \citep[e.g.][]{polarization_of_clouds}. 
The orbits with $i < 52^\circ$ do not include the rainbow location
and we distribute the observation epochs evenly along those orbits,
in steps of $45^\circ$ around the star (see 
Fig.~\ref{fig_orbital_positions})
to allow the observer to sample all parts of the surface. 
For the orbits with $i \geq 52^\circ$, we choose two 
observation locations at $\alpha=38^\circ$ and two at 
$\alpha = 142^\circ$, and the other four locations are 
distributed evenly along the remaining orbit.
The location of each observational epoch with the corresponding
value of $\alpha$ is listed in Table~\ref{tab_orbital_positions}.
Since changing the orbital eccentricity and radius alters 
only the magnitude and timing of the phase curves, 
we use circular orbits with $r=1$ and $F_0=1$.

An exoplanet's rotational period and axis of rotation modulate the 
observed light curves and must therefore be retrieved to map an 
exoplanet's surface \citep{frequency_modulation}. 
We use a set of rotation axes that are homogeneously distributed 
across a unit sphere's surface defined in the Cartesian 
coordinate system as $(x,y,z)$ (see Fig.~\ref{fig_orbital_positions}), 
as this prevents the neural networks from learning a bias.  
So that the neural networks cannot resolve the degeneracy described in 
Sect.~\ref{subsec:degeneracy}, the degenerate counterpart of each 
rotation axis is also included in the set.
To achieve this, the first $32$ points from a $64$-point Fibonacci 
spiral \citep[see][]{fibonacci} as well as their degenerate counterparts 
(found by multiplying element-wise by $(-1,-1,1)$) create a set of $64$ 
axes.

At each of the eight observation locations, the model planet  
is observed eight times as is rotates about its axis,
for a total of $64$ observations per orbit (we use a constant 
phase angle at each location, neglecting the orbital motion).
During each observation, the planet is assumed to have a fixed
orientation. We thus do not include rotation during the integration
period (see also recommendation vi in Sect.~\ref{sect_conclusions}).
This approach reduces memory requirements for storing the phase
curves, simplifies the network architecture, and increases 
neuron gradients, 
leading to more efficient training of the neural network.

Figure~\ref{fig_earth_curves} shows the total and
polarized fluxes of the Earth-like
planet shown in Fig.~\ref{fig_earth_with_clouds} in an edge-on orbit
($i=90^\circ$) and with a rotation axis angle of 90$^\circ$. 
Curves are shown both for
a Lambertian and a bidirectionally reflecting model planet. 
Since the Earth-like planet is largely covered with ocean 
and clouds, the Lambertian and bidirectional curves differ 
significantly in magnitude, particularly when 
$\alpha=38^\circ$, and when $\alpha=142^\circ$
(see Table~\ref{tab_orbital_positions}).
This was to be expected from the reflection patterns of the spatially 
resolved planets shown in Fig.~\ref{fig_resolved_fluxes}.

%--------------------------------------------------------------------------
% Figure 8:
%--------------------------------------------------------------------------
\begin{figure}[ht!]
\centering
\includegraphics[width=87mm]{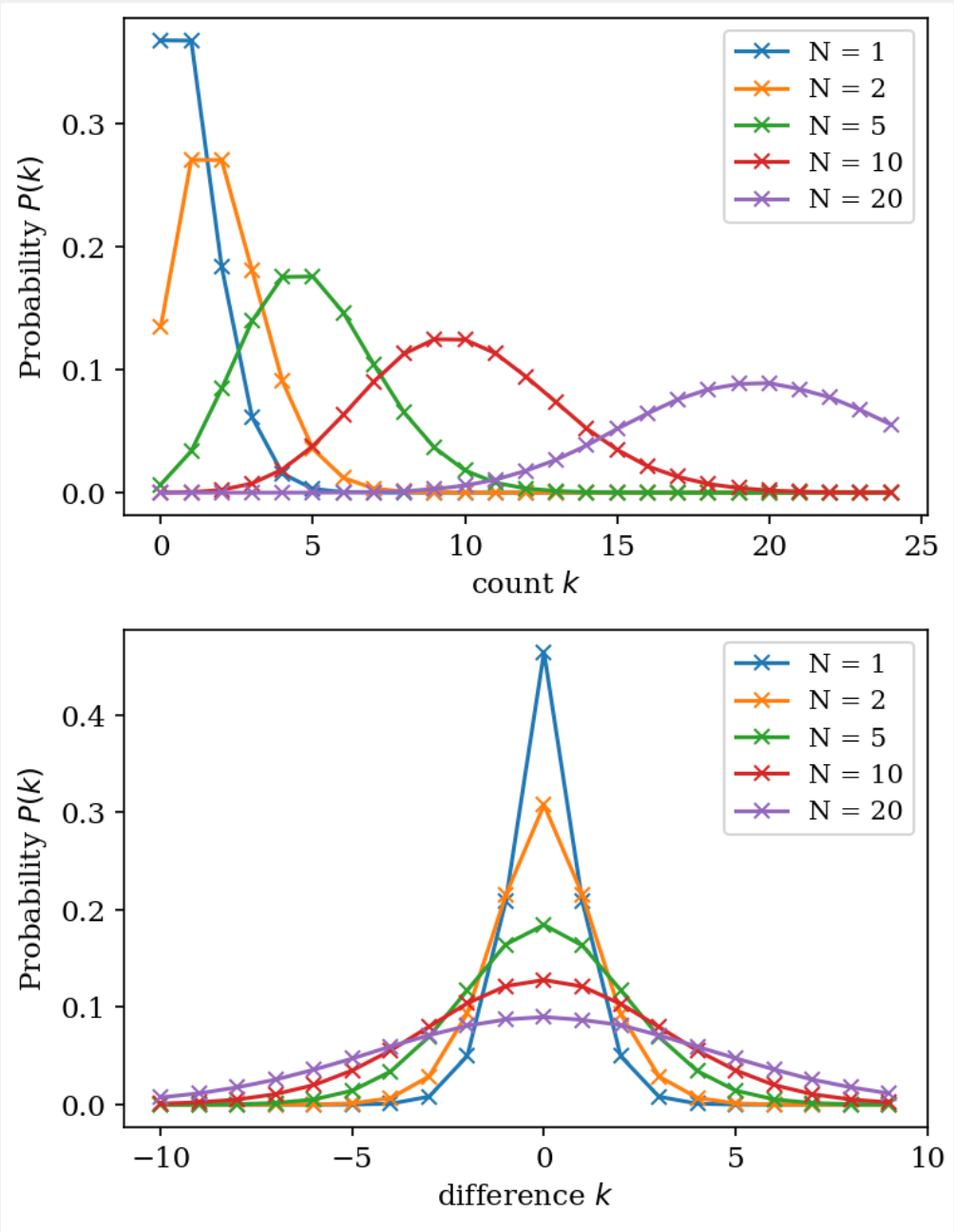}
\caption{The Poisson distributions with mean $N$ (top) give
        the probability for $k$ photons being received from a light 
        source (such as a reflecting planet). 
        The Skellam distributions (bottom) give the probability 
        for the difference of two independent Poisson-distributed variables, describing the polarized fluxes $Q$ and $U$.}
\label{fig_prob_distributions}
\end{figure}
%--------------------------------------------------------------------------
%--------------------------------------------------------------------------
% Figure 9:
%--------------------------------------------------------------------------
\begin{figure}[h!]
\centering
\includegraphics[width=87mm]{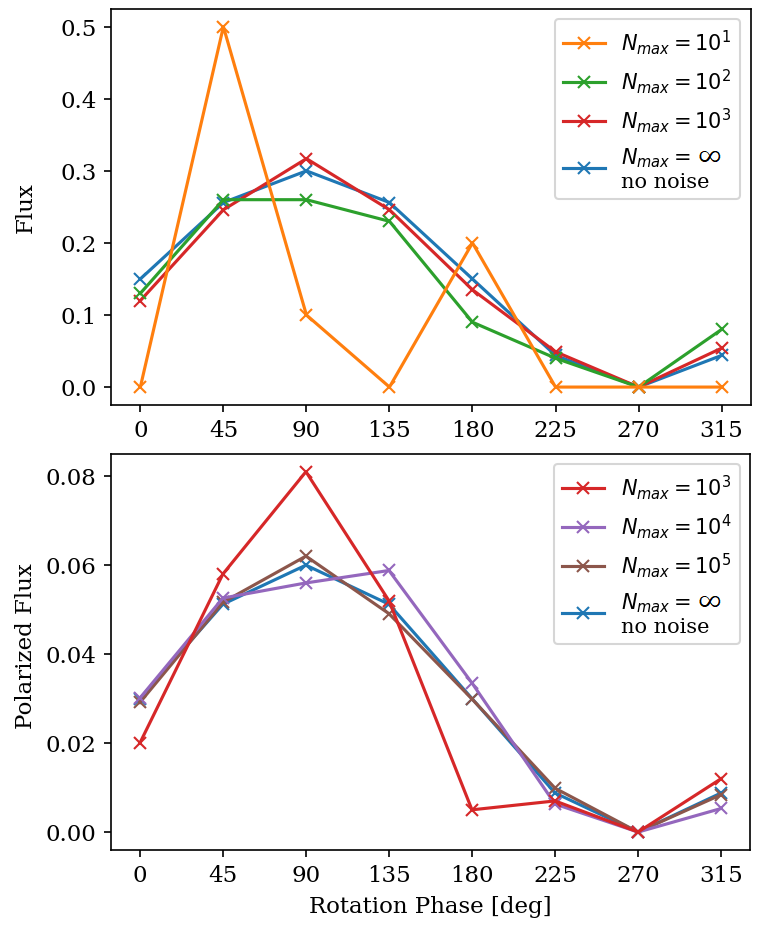}
\caption{Total (top) and polarized (bottom) flux curves at various noise 
         levels. The number of photons that represents a total flux equal 
         to $1.0$ (a white Lambertian planet at $\alpha=0^\circ$) is
         $N_\text{max}$; the range of $N_\text{max}$ used for the 
         polarized flux differs from that used for the total flux,
         because the former is generally smaller than the 
         latter (cf.\ Fig.~\ref{fig_earth_curves}) and thus more 
         sensitive to noise. The curves labeled `no noise' 
         ($N_{\text{max}}=\infty$) are sine functions with similar
         amplitudes as the curves in Fig.~\ref{fig_earth_curves}.}
\label{fig_curves_with_noise}
\end{figure}
%--------------------------------------------------------------------------

%%%%%%%%%%%%%%%%%%%%%%%%%%%%%%%%%%%%%%%%%%%%%%%%%%%%%%%%%%%%%%%%%%%%%%%%%%
\subsection{Photon noise}
\label{sec:noise}

Because the light fluxes reflected by exoplanets will be 
very low, the measurable signals will have relatively strong noise levels.
Most exocartography research uses
Gaussian noise \citep[e.g.\ ][]{nn_cartography}, which
can be compared to instrumental noise. In that case, however, 
the noise level is not adjusted to the magnitude of the flux, 
which can give negative values when the flux is low.
Instead we use Poisson noise (i.e.\ photon noise). 
%to describe the 
%noise that arises from the particle nature of light.
To calculate the noise in the total and polarized fluxes,
we write the observed fluxes $I_\text{o}$, $Q_\text{o}$, 
and $U_\text{o}$ as functions of the linearly polarized fluxes 
$I_x$, with $x$ the angle of the optical axis of the 
linear polarizer through which the flux was measured
\citep[see Eq.~(1.6) of][]{scattering_in_atmospheres}:
\begin{equation}
    I_\text{o} = I_{0^\circ} + I_{90^\circ}, \hspace{5mm} Q_\text{o} = I_{0^\circ} - I_{90^\circ}, \hspace{5mm} U_\text{o} = I_{45^\circ} - I_{135^\circ},
\label{eq:I_and_Q}    
\end{equation}
where %$I_{x^\circ}$,
$I_x$ can be derived from the numerically simulated
noise-free fluxes $I$, $Q$, and $U$ using Eq.\ 1.5 from
\citet{scattering_in_atmospheres} (and ignoring circularly polarized flux $V$):
\begin{equation}
    I_x = \tfrac{1}{2} (I + Q \cos 2x + U \sin 2x).
\end{equation}
For each noise-free flux $I_x$, we then compute the 
corresponding number of photons by multiplication with $N_\text{max}$, 
the number of 
photons that would be received from a white Lambertian planet at full 
phase ($\alpha=0^\circ$) without noise. 
Then the noisy $I_x$ are obtained by drawing from Poisson distributions,
and the observed, noisy $I_\text{o}$, $Q_\text{o}$, and $U_\text{o}$
are computed using Eq.~(\ref{eq:I_and_Q}).
The result of adding two Poisson-distributed variables with averages 
$N=\mu_1$ and $N=\mu_2$ is a Poisson-distributed variable with 
$N=\mu_1+\mu_2$, so $I_\text{o}$ is also
described by a Poisson distribution. 
When subtracting two independent Poisson-distributed variables, 
such as for computing $Q_\text{o}$ and $U_\text{o}$,
one obtains a so-called Skellam distribution with mean $\mu_1-\mu_2$ and 
variance $\mu_1+\mu_2$. 
The distributions we use are shown in Fig.~\ref{fig_prob_distributions} 
for $\mu_1=\mu_2$.

The effect of noise on flux curves can be seen in
Fig.~\ref{fig_curves_with_noise}, where we show simulated total
and polarized flux curves for eight rotational phases (thus at 
a single orbital position) assuming different photon numbers and thus
different noise levels. Because polarized fluxes are generally much
smaller than total fluxes (see Fig.~\ref{fig_earth_curves}), 
they are more sensitive to photon noise. This is 
particularly obvious when comparing the curves for $N_{\text{max}}=10^3$ in 
the two graphs of Fig.~\ref{fig_curves_with_noise}: while for the 
total flux this curve almost coincides with the $N_{\text{max}}=\infty$ 
(`no noise') curve,
for the polarized flux the difference is large except at the 
highest flux level ($N_{\text{max}}=10^5$). 
Indeed, for the polarized flux to be 
reliably observed, we would need $N_{\text{max}} \approx 2\cdot 10^4$.

The number of reflected photons received by a telescope from
a white Lambertian planet at $\alpha=0^\circ$ 
can be derived starting with Eq.~(\ref{eq_reflectedstokesvector})
and using $A_{\rm G} \vec{R}_{11}= \frac{2}{3}$ for 
the reflection by the Lambertian planet
\citep[see e.g.][]{stam_2006,1957lssp.book.....V}:
\begin{equation}
    N_\text{max} = 
           \frac{2}{3} \hspace*{0.1cm}
           \frac{r^2}{D^2} \hspace*{0.1cm} 
           \frac{\dot{N}}{4\pi d^2} \hspace*{0.1cm} 
           t_\text{integ.} 
           \hspace*{0.1cm}
           \pi r_\text{tel.}^2.
\label{eq:Nmax}
\end{equation}
Here, $\dot{N}$ is star's photon emission rate  
across the spectral
band under consideration, $t_\text{integ.}$ is the integration time, 
$D$ is the distance between the telescope and the star 
system, and $r_\text{tel.}$, $r$, and 
$d$ are the respective radii of the telescope's 
primary mirror, the planet and the planetary orbit.

In the following, we assume that the planet is Earth-sized and
orbits its solar-type star at a distance of $1~\text{au}$. The 
integration time $t_\text{integ.}$ is 3 hours (24/8 hours/observations),
and the spectral bandwidth is $50~\text{nm}$.
The stellar flux across each spectral band is computed assuming 
black-body radiation and a stellar radius and effective temperature of, 
respectively, $695{\small,}700~\text{km}$ and $5772~\text{K}$,
similar to solar values.
Figure~\ref{fig_number_of_photons} shows that the number of photons
$N_\text{max}$ reflected by a white, Lambertian reflecting
planet in the Alpha Centauri system would range from $10^5$ to $10^6$.
As we will show in Sect.~\ref{sec:map_retrieval_accuracies}, 
$N_\text{max}$ should be $10^4$ or more for accurate retrievals. 
According to Eq.~(\ref{eq:Nmax}), this should be achievable
for systems up to distances of $20$ lightyears with the Nancy-Roman
Space telescope (in combination with a star shade) 
and $30$ lightyears with the HabEx telescope.

%We imagine that the actual observation campaign consist of the following steps.}
%\begin{enumerate}[i]
%    \item 
%    Determine the orbital elements of the Kepler orbit 
%    and the period from the (relative) positions of the planet at the observation points. \citet{Feng_2019} describe a comprehensive method to obtain these.
%    \item
%    Determine the diurnal period from light curves at (one or more) observation points, and the phase shift between the points \citep[see for example][]{Visser2015}.
%    Close-in planets can be tidally locked or in a spin-orbit resonance. These need to be treated as special cases that we do not consider in this paper.
%    \item
%    Update the parameters, and correct 
%    for the diurnal phase shift due to the light-travel (R{\o}mer) delay, at the different observation points.
%    \item
%    Retrieve the rotation axis as explained in Sect.\ \ref{sect_results_rotation_axis}.
%    \item
%    Retrieve the surface map. This is demonstrated in Sect.\ \ref{sect_surface_type_maps}.
%\end{enumerate}

%%%%%%%%%%%%%%%%%%%%%%%%%%%%%%%%%%%%%%%%%%%%%%%%%%%%%%%%%%%%%%%%%%%%%%%%%%
\section{Method: neural networks}
\label{sect_neural_networks}

%%%%%%%%%%%%%%%%%%%%%%%%%%%%%%%%%%%%%%%%%%%%%%%%%%%%%%%%%%%%%%%%%%%%%%%%%%
\subsection{Training our neural networks}
\label{sect_training_the_network}

Our neural networks are trained by feeding them data 
of model planets and by comparing the networks' output to each
model planet's true rotation axis orientation and surface map. 
The training data consists of numerically simulated observations
of model planets with various surface and cloud maps, 
inclinations and rotation axis angles. These input values 
are multiplied by neuron weights that are adjusted by an optimization 
algorithm minimizing the loss, defined by the difference between the 
desired output and the network's output. This process is repeated in 
epochs until the optimization algorithm no longer increases the network's
accuracy. The latter is established by evaluating the loss of the 
network when applied to validation planets, i.e.\ model planets that
were not included in the training data set.

The large parameter space requires a large training data set. 
To reduce computation and memory requirements, we define a set of 
possibilities for each of the parameters. 
The set size of each of the four parameters is 
listed in Table~\ref{tab_parameter_set}, resulting in a total 
of 44.8~billion unique combinations. We rapidly 
create 4~million model planets by randomly drawing combinations
from the sets.
For these planets we then compute flux curves that is the input 
to the network.
One tenth of the planets and light curves are used for validation, 
to evaluate the accuracy of the neural networks.

%-------------------------------------------------------------------------
% Table 2:
%-------------------------------------------------------------------------
\begin{table}[h!]
\centering
\caption{Number of combinations for which we created model planets.}
\begin{tabular}{lr}
\hline\hline
Parameter & \# combinations \\ \hline
% Surface Types & $4$ \\ %\hline
Inclinations & $7$ \\ %\hline
Rotation Axes & $64$ \\ %\hline
%Orbital Locations & $8$ \\ %\hline
%Phases of Rotation & $8$ \\ %\hline
%Facets & $1000$ \\ %\hline
%Wavelengths & $6$ \\ %\hline
%Stokes Parameters & $3$ \\ 
Cloud maps & 10\small{,}000 \\ 
Surface maps & 10\small{,}000 \\ \hline     
\end{tabular}
\tablefoot{These combinations yield 44.8~billion planets
           from which we select 4~million for the training 
           data set and for validation.
}
\label{tab_parameter_set}
\end{table}
%-------------------------------------------------------------------------

%%%%%%%%%%%%%%%%%%%%%%%%%%%%%%%%%%%%%%%%%%%%%%%%%%%%%%%%%%%%%%%%%%%%%%%%%%
\subsection{Periodic convolutions}
\label{subsec:periodic_convolutions}

\begin{figure}[t!]
\centering
\includegraphics[width = 0.8\linewidth]{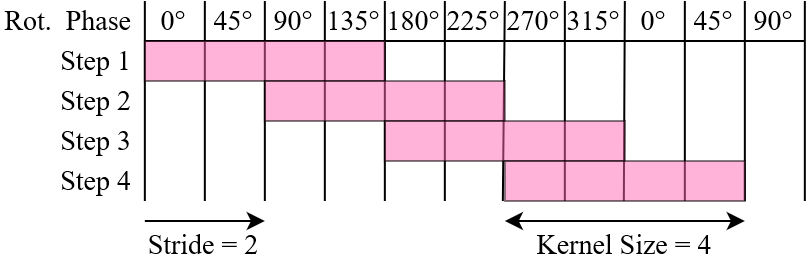}
\caption{Periodic convolution with a stride of 2 and kernel size $N$ of 4. 
         Since the rotational phases are pseudo-periodic, the first $N-1$ 
         values are appended to the end of the light curves before the 
         size $N$ kernel slides over to preserve dimensions. 
}
\label{fig:periodic_convolutions}
\end{figure}

Recently, 1-D convolutions have been used to process time-series 
in neural networks, achieving state-of-the-art results in fields such as 
biomedical data classification and early diagnosis, and anomaly detection 
in power electronics and electrical motor fault detection 
\citep{1D_convolutions}. 
Our neural network architecture uses such 1-D convolutions, 
adjusted to take advantage of the periodic nature of the light curves.
Since the model planet's prime meridian (the line of $0^\circ$ longitude) 
is arbitrary, the relationship between the rotational phases of,
e.g., $315^\circ$ and $0^\circ$ should be equivalent to that
between e.g.\ $45^\circ$ and $90^\circ$.
To prevent that such combinations are interpreted as being distinct
by the neural network, the first $N-1$ rotational phases are appended 
to the end of each light curve before the $1\times N$ convolutional 
kernel slides over the light curve, as shown
in Fig.~\ref{fig:periodic_convolutions} for a kernel size $N$ of 4.
An additional advantage is that the convolution does not change 
the dimensions of the light curve. There is thus no need for 
zero-padding, a common method for preserving data dimensions by 
adding a border of zeros around the actual data, which has the 
disadvantage of increasing susceptibility 
to spatial bias, as shown by \citet{mind_the_pad}.

%---------------------------------------------------------------------------
% Figure 10:
%---------------------------------------------------------------------------
%---------------------------------------------------------------------------

%---------------------------------------------------------------------------
% Figure 11:
%---------------------------------------------------------------------------
\begin{figure}[b!]
\centering
\includegraphics[width=.3\linewidth]{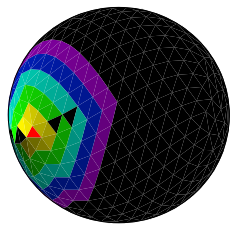}
\caption{An example of a spherical convolution with five rings.
         The method developed by \citet{spherical_convolutions} 
         (originally for HEALPix) is adapted to the Fibonacci sphere 
         with the improvement that instead of zero-padding, 
         the periodic nature of rings is used. Some facets inside 
         the rings are black because they are not included in the kernel,
         since the number of facets in each respective ring must match 
         for all kernel locations on the sphere.}
\label{fig:spherical_convolutions}
\end{figure}
%---------------------------------------------------------------------------

To down-sample the data dimensions, a stride greater than~$1$ can be 
used. However, the kernel size should be a multiple of the stride 
to avoid some rotational phases being sampled more than others. 
We use a kernel size of $1\times 4$ and a stride of $2$ 
in the final architecture and found that the loss of the neural network 
decreases by 10\% when using these periodic convolutions rather than 
conventional 1-D convolutions. 

%%%%%%%%%%%%%%%%%%%%%%%%%%%%%%%%%%%%%%%%%%%%%%%%%%%%%%%%%%%%%%%%%%%%%%%%%%
\subsection{Spherical convolutions}
\label{subsec:spherical_convolutions}

Originally developed for image recognition \citep{lecun1989}, 
2-D convolutions can also be used for 2-D image generation, for 
example when generating faces \citep{stylegan2} or for semantic
segmentation, where the output of the neural network is an image with 
each pixel belonging to a specific class \citep{semantic_segmentation}. 
\citet{nn_cartography} demonstrated that convolutions on the surface 
of a sphere can be used to regularize retrieved exoplanet maps. 
They use the spherical convolution algorithm developed by
\citet{spherical_convolutions} combined with ReLU activation functions 
to regularize a retrieved planet map on a HEALPix pixelization scheme 
\citep{gorski_2005}.

Since we do not use HEALPix but a Fibonacci sphere, we have adapted 
the spherical convolution algorithm by \citet{spherical_convolutions}. 
A spherical convolution takes advantage of 
the optimization that has been achieved for 1-D convolutions by 
expanding the 1-D list of facets on the sphere's surface such that 
each facet is followed by its $N$ surrounding facets. Then a kernel 
with size and stride equal to $N + 1$
convolves each facet with its surrounding facets. 
To achieve this for a Fibonacci sphere, we first identify each facet's
surrounding facets. These are then ordered in clockwise direction by 
calculating the clockwise angle from the $z$-direction to each 
surrounding facet's center. 
More rings can be added by taking the surrounding facets of the 
inner ring. A problem that arises here is that the rings 
around the facets have different sizes, while the 1-D kernel 
can only have one size. To solve this problem,
\citet{spherical_convolutions} use zero-padding, 
which can cause unwanted artefacts \citep{mind_the_pad}. 
To avoid these artefacts, 
we instead use the periodicity of the rings and add the first 
value to the end of the series. 
Then we compute the mean integer length of the rings surrounding 
each facet, and if the number of facets in a given ring differs from the 
mean value, facets are either repeated or taken out in an evenly spaced
manner. This can yield gaps in the rings as shown in 
Fig.~\ref{fig:spherical_convolutions}, while in other places there can
be double counts, causing spatial bias with minimal effects on the final 
maps.

%%%%%%%%%%%%%%%%%%%%%%%%%%%%%%%%%%%%%%%%%%%%%%%%%%%%%%%%%%%%%%%%%%%%%%%%%%%%%%
%--------------------------------------------------------------------------
% Figure 12
%--------------------------------------------------------------------------
\begin{figure*}[t!]
\centering
\includegraphics[width = \textwidth]{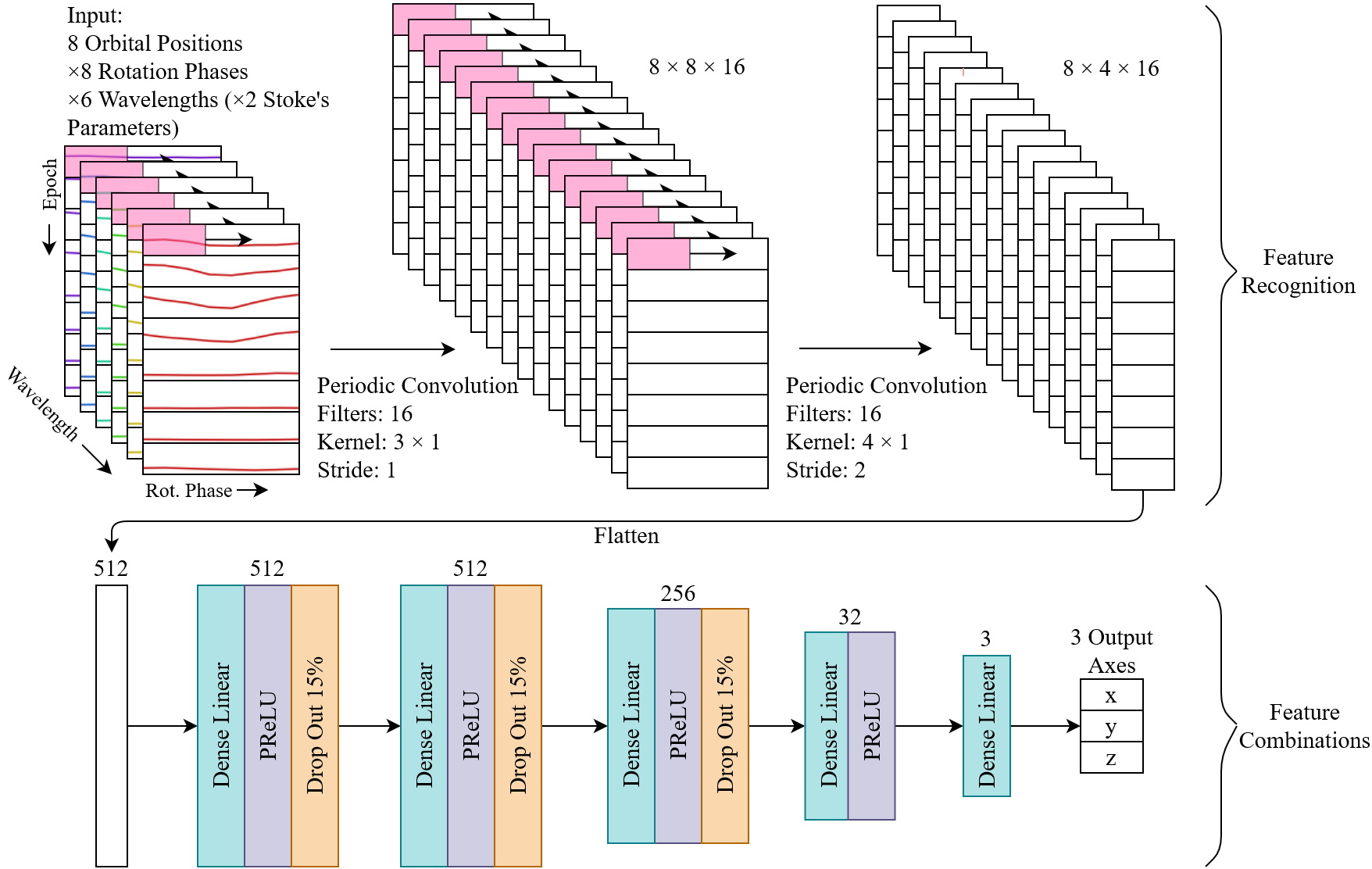}
    \caption{The network architecture to retrieve rotation axes. 
             The feature recognition part uses periodic convolutions 
             and the feature combinations part consists of densely 
             connected layers with PReLU activation functions and 
             dropout layers to prevent over-fitting (the number of nodes
             is indicated above the lower layers). When using 
             polarization, the input shape is 
             $8\times 8\times 12$, otherwise $8\times 8\times 6$. 
             The periodic convolutions maintain the 1st and 2nd data 
             dimensions since the first $N - 1$ values along the 
             rotation axis are appended to the end before convolution. 
             The number of filters determines the 3rd dimension of the 
             output. The number of trainable parameters is 
             $667\small{,}907$ (with polarization) and $667\small{,}619$
             (without polarization).}
\label{fig:final_nn}
\end{figure*}
%--------------------------------------------------------------------------

%--------------------------------------------------------------------------
% Figure 13
%--------------------------------------------------------------------------
\begin{figure}[htb!]
\includegraphics[width=0.6\linewidth]{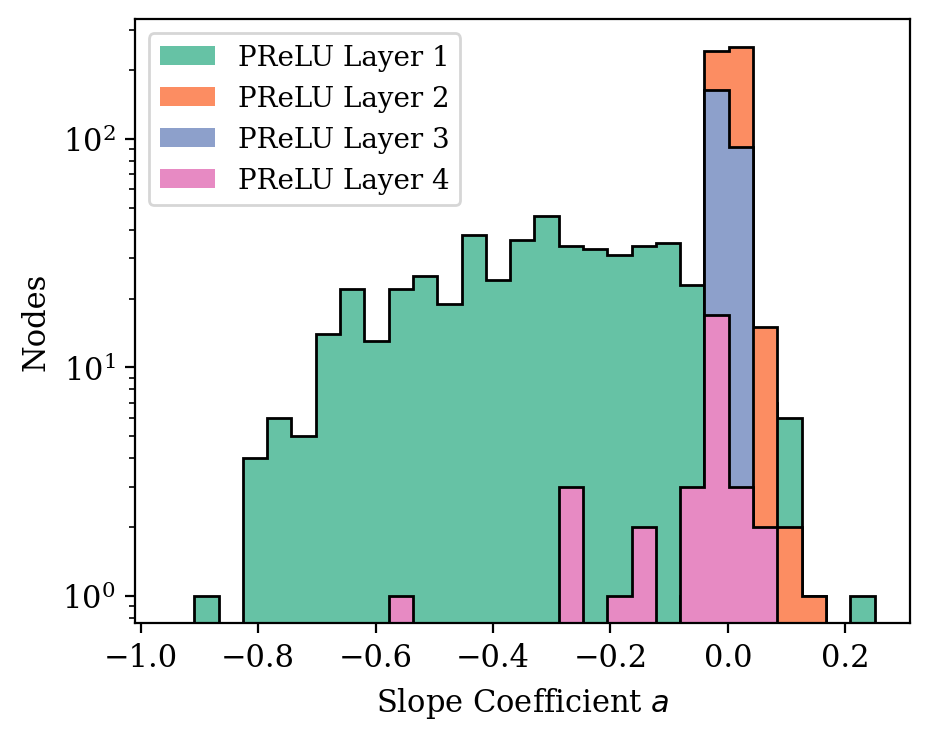}
\caption{PReLU slope coefficients after training (plotted with a log scale).
         See Fig.~\ref{fig:final_nn} for the position of each layer in the
         neural network. These coefficients were retrieved from a neural 
         network trained on planets in an edge-on orbit with bidirectional
         light curves and including polarization.}
\label{fig:slope_coefficients}
\end{figure}

%--------------------------------------------------------------------------

\section{Results: retrieval of the rotation axis}
\label{sect_results_rotation_axis}

The first step for retrieving planet maps using reflected fluxes
is to estimate the orientation of the planet's rotation axis.
\citet{fujii_2012} find the axis orientation
by maximizing the overlap between the measured signals
and dominant left-singular vectors.
\citet{frequency_modulation} also showed that retrieval of the axis
is possible from measurements of the frequency modulation of the 
planetary signal over a complete orbit. 
Because we take eight discrete observational periods of a planet 
along its orbit instead of (near) continuous monitoring, we decided
to use a convolutional neural network approach
with as output of the network the three Cartesian 
coordinates of the planet's rotation axis.

%%%%%%%%%%%%%%%%%%%%%%%%%%%%%%%%%%%%%%%%%%%%%%%%%%%%%%%%%%%%%%%%%%%%%%%%%%%
\subsection{Degeneracy in the signal for the edge-on case}
\label{subsec:degeneracy}

The retrieval of the rotation axis is problematic when the orbital
inclination angle $i$ is close to $90^\circ$, because then every
configuration has a mirror configuration with the same light curve 
upon reflection in the orbital plane, as the orbit and the sense 
of rotation remain the same. 
However, this leaves the planet map mirrored with respect to the 
equator and the $x$ and $y$ components of the rotation axis change sign. 
In the reference system shown in Fig.~\ref{fig_orbital_positions}, 
this would be expressed as an element-wise multiplication by $(-1,-1,1)$.
Because the two configurations are inherently different, while the 
observable light curves are identical, the neural network should not
be able to decide on the actual configuration.
We identified the degeneracy by trial-and-error and confirmed by testing 
that the light curves are the same for several different mirror pairs. 

%%%%%%%%%%%%%%%%%%%%%%%%%%%%%%%%%%%%%%%%%%%%%%%%%%%%%%%%%%%%%%%%%%%%%%%%%%%
\subsection{Architecture}
\label{sec:final_nn}

We draw inspiration for our neural network architecture, shown in 
Fig.~\ref{fig:final_nn}, from other convolutional signal processing 
neural networks \citep[see, for example,][]{speech_emotion_recognition}. 
We split the network into a `feature recognition' part with convolutional 
layers followed by a `feature combination' part with dense linear layers. 
Note that the network must be retrained for each new orbital inclination
angle~$i$.

The input dimensions to the neural network are either $8\times 8\times 6$ 
without polarization ($8$ orbital locations, $8$ rotational phases, $6$ 
wavelengths) or $8\times 8\times 12$ with polarization 
(Stokes parameters $I$ and $Q$ are included, which doubles the final 
dimension; parameters $U$ and $V$ have negligible magnitudes 
and are not included in the retrievals). 
The network's feature recognition part consists of two periodic 
convolutional layers, each with $16$ filters. 
The first layer, with a kernel size of $3$, increases the dimensions 
of the input to $8\times 8\times 16$, as shown in Fig.~\ref{fig:final_nn}. 
The final dimension increases to $16$ since each of the filters, 
which convolve the $6$ wavelengths and $8$ rotational phases, 
has its own output.
To halve the size of the data, the second periodic convolution skips 
half of the rotational phase steps
($\text{stride}=2$) with a kernel size of $4$, 
leading to an $8\times 4\times 16$ output, as shown in 
Fig.~\ref{fig:periodic_convolutions}. The output of the last
convolutional layer is then flattened and passed onto the network's 
feature combination part.

The data is then fed through five dense, fully connected layers with 
linear activation functions and a bias. These layers have $512$, $512$, 
$256$, $32$ and $3$ nodes, respectively. The first four layers are 
followed by PReLU activation functions to introduce non-linearity 
\citep{prelu}:
\begin{equation}
\label{eq:prelu}
f\left(y_{k}\right) = 
\left\{
\begin{array}{ll}
y_{k} & \text{if} \quad y_{k} > 0 \\
a_{k} \hspace*{0.05cm} y_{k} & \text{if} \quad  y_{k} \leq 0
\end{array}
\right.
,
\end{equation}
where the slope coefficient $a_i$ causes non-linearity due to 
the change in gradient at $y_i = 0$ and is actively learned and 
manually inspected (see Fig.~\ref{fig:slope_coefficients}). 
To prevent over-fitting, the first three layers are followed by 
dropout regularization layers with a dropout rate of $15\%$. 
These dropout layers randomly set $15\%$ of the layer's output to $0$, 
preventing the neural network from over-fitting to peculiarities 
of the training data. The network has a total of $667,907$ or $667,619$
trainable parameters, depending on whether polarization is included 
or not, respectively. The parameters that we train are the weights 
of the convolutional kernels, linear nodes and slope coefficients 
of the PReLU layers.

%--------------------------------------------------------------------------
% Figure 14
%--------------------------------------------------------------------------
\begin{figure*}[ht!]
\centering
\includegraphics[width = \textwidth]{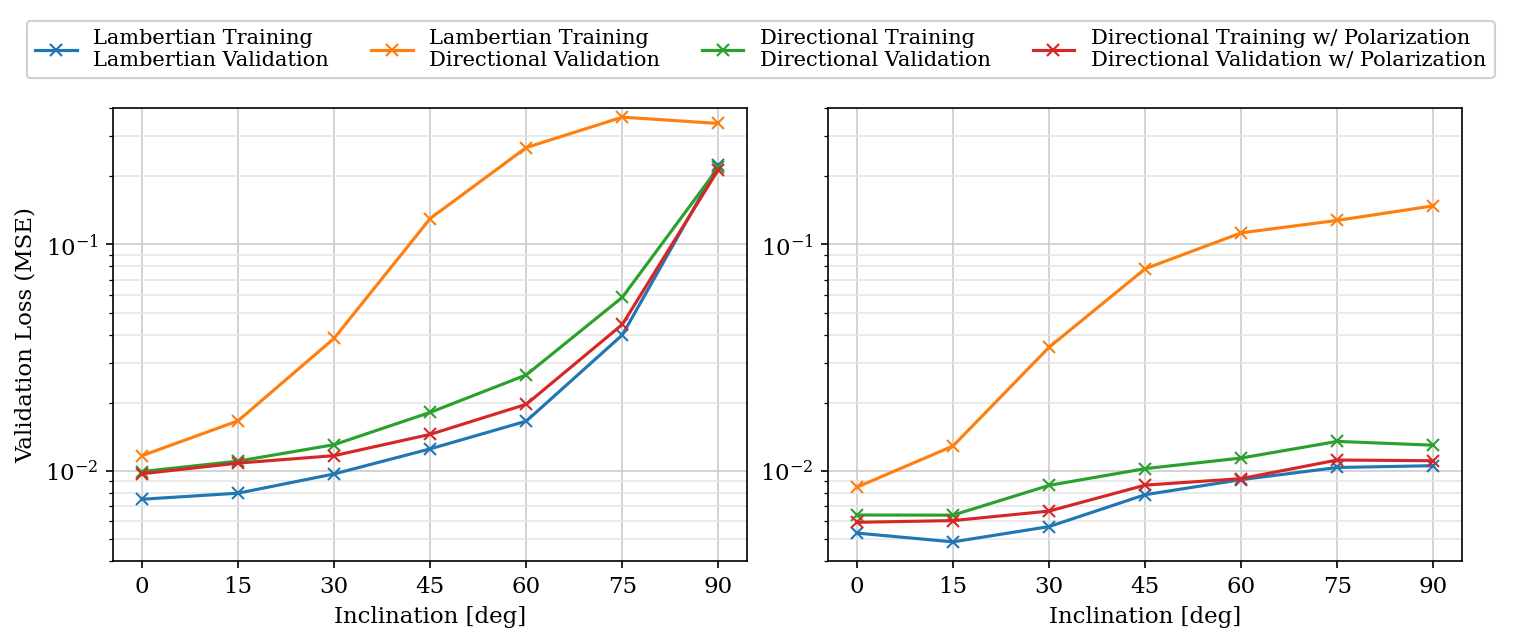}
\caption{The rotation axis retrieval accuracy as a function of 
          orbital inclination angle $i$ without (left) and with 
          the constraint $y\geq 0$ (right).
          Blue:  training with Lambertian reflection 
          and retrieval with Lambertian reflection;
          Orange: training with bidirectional reflection 
          and retrieval with Lambertian reflection;
          Green: training with bidirectional reflection 
          and retrieval with bidirectional reflection, both
          without polarization; 
          Red: the same, but both with polarization.
}
\label{fig:incl_losses}
\end{figure*}
%--------------------------------------------------------------------------

%--------------------------------------------------------------------------
% Figure 15
%--------------------------------------------------------------------------
\begin{figure*}[hb!]
\centering
\includegraphics[width = \textwidth]{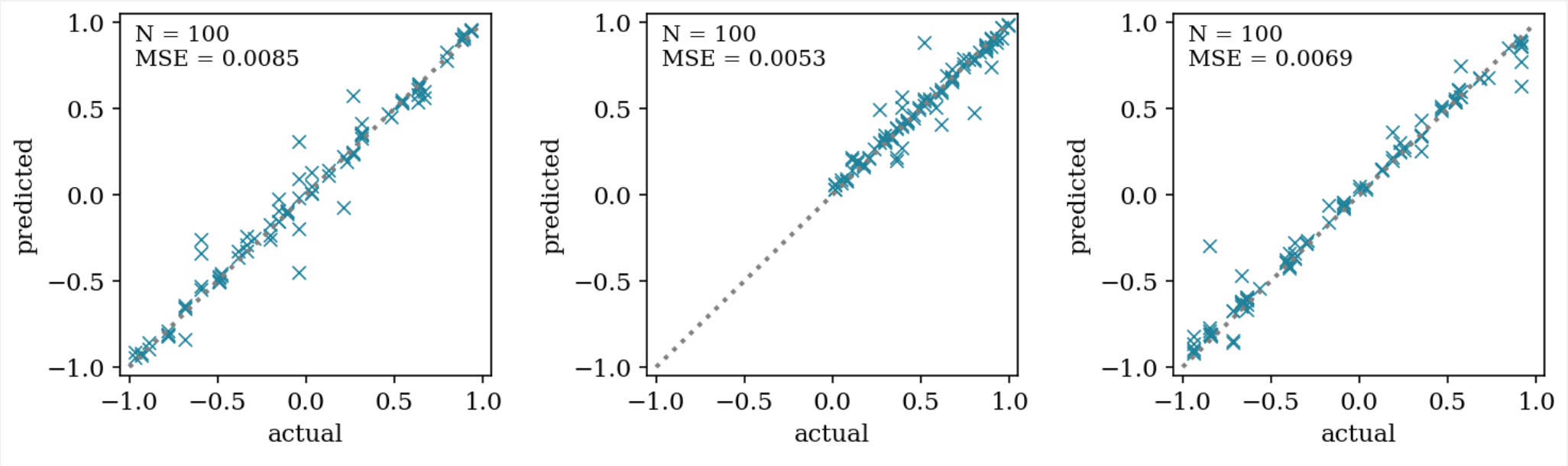}
\caption{The accuracy of the retrieval of the rotation axis direction
         for $N=100$ model planets in edge-on orbits: the $x$-coordinate 
         (left), $y$-coordinate (middle), and $z$-coordinate (right).
         The degeneracy has been mitigated with the constraint
         $y \geq 0$.}
\label{fig:side_on_losses_constrained}
\end{figure*}
%--------------------------------------------------------------------------

%%%%%%%%%%%%%%%%%%%%%%%%%%%%%%%%%%%%%%%%%%%%%%%%%%%%%%%%%%%%%%%%%%%%%%%%%%%
\subsection{Training the axis retrieval network}

When the neural network is trained on all planets with a 
given inclination angle $i$, the available number of light
curves is $4\small{,}000\small{,}000 / 7 \approx 570\small{,}000$. 
In order to mitigate the degeneracy (Sect.~\ref{subsec:degeneracy}), 
the network can only be trained on all planets with a rotation axis 
with $y \geq$ 0, in which case the number of curves is halved.  
Note that we use $90\%$ of the curves for training and $10\%$ 
for validation. To prevent over-fitting, we stop with training when 
the validation losses have not decreased for two consecutive epochs.

The neural network is created and trained using the Keras Python 
package.\footnote{\url{https://keras.io/}} Since retrieving the 
rotation axis is a regression problem, 
we chose the mean squared error (MSE) as the loss that the network 
should minimize. We found that for small batch sizes ($16$ to $64$ 
planets fed to the neural network together), the training loss can 
increase after roughly ten epochs. 
This appears to indicate that the optimization algorithm
\citep{adam} overshoots the local minimum and adjusts to peculiarities 
in the small batch, rather than finding the global optimum.
This problem was solved by increasing the batch size to $256$.

Figure~\ref{fig:slope_coefficients} shows the slope coefficients $a$ 
(see Eq.~(\ref{eq:prelu})) after training for the edge-on case. 
The coefficients of PReLU layers 2, 3, and 4 are distributed around zero.
Because they approximate a classical ReLU function with $a = 0$, 
we tested replacing them with ReLU layers, but this did not lead to 
better results as the validation losses of this modified network were
larger; apparently the non-zero $a$ values in these PReLU layers 
do add value. 
PReLU layer 1 is not centered around zero but has 
a mean $a$ of $-0.32$, with a standard deviation that is roughly 
one order of magnitude larger than that for the other layers. 
The negative coefficients of the majority of the nodes in layer~1 
mean that the PReLU function is not one-to-one.

%%%%%%%%%%%%%%%%%%%%%%%%%%%%%%%%%%%%%%%%%%%%%%%%%%%%%%%%%%%%%%%%%%%%%%%%%%%
\subsection{Numerical artefacts}
\label{sec:numerical_artefacts}

When the network is trained with light curves with very little noise, 
it can distinguish between the two degenerate cases addressed in 
Sect.~\ref{subsec:degeneracy} and correctly estimate the $x$ and 
$y$ coordinates of the planet's rotation axis. 
Because this information should actually not be present in the light 
curves, this shows that the network uses numerical artefacts in the 
retrieval. 
For example, the network might have learned to recognize the facet 
scheme, which is not symmetric under the reflection in the plane.

By increasing the noise levels, such numerical artefacts are suppressed. 
We have determined by trial and error that a noise level at photon numbers $N_\text{max} \approx 10^4$ ensures that the network is unable to distinguish the degenerate cases, 
as it should be.

%%%%%%%%%%%%%%%%%%%%%%%%%%%%%%%%%%%%%%%%%%%%%%%%%%%%%%%%%%%%%%%%%%%%%%%%%%%
\subsection{Retrieval accuracy for near edge-on observations}
\label{sec:rot_axis_retrieval_accuracy}

The left panel in Fig.~\ref{fig:incl_losses} shows the retrieval 
accuracy of the neural 
network as a function of the orbital inclination angle $i$.
The loss increases with $i$ up to $\text{MSE}=0.22$ for $i=90^\circ$,
except for the network that was trained with Lambertian data
but that was presented directional validation. 
Here, the degeneracy in the geometry happens:
the network cannot determine the signs of the $x$ and $y$ components 
of the rotation axes. In that case the retrieved values are near zero 
and therefore
\begin{equation}
\text{MSE} = \tfrac{1}{3}\langle\Delta x^2+\Delta y^2 + \Delta z^2\rangle \approx 
%\tfrac{1}{3}\langle\Delta x^2+\Delta y^2\rangle =
\tfrac{1}{3}\langle x^2+y^2\rangle = \tfrac{2}{9}
.
\end{equation}

The degeneracy for the near edge-on observation can be mitigated by 
constraining the rotation axes to $y \geq 0$ and by training 
the neural network on these planets only. The neural network is able to 
approximate the rotation axis in this half of the search space, 
as shown in Fig.~\ref{fig:side_on_losses_constrained}.
With decreasing inclination angle,  the two cases $y>0$ and $y<0$ are
increasingly distinguishable, and the use of a constraint becomes invalid,
as by then giving away the correct sign of $y$, one obviously 
artificially reduces the MSE.

%%%%%%%%%%%%%%%%%%%%%%%%%%%%%%%%%%%%%%%%%%%%%%%%%%%%%%%%%%%%%%%%%%%%%%%%%%%
\subsection{Retrieval accuracy at other inclination angles}
\label{subsec:without_degeneracy}

The network is trained three times, for three different reflection models. 
First, we trained it with light curves computed assuming Lambertian 
reflection. 
The second and third training sessions are with light curves computed 
using bidirectional reflection with and without polarization. 
Note that all losses discussed in this section are validation losses, 
and thus the network's losses when it is applied to the 
validation planets that it was not trained with. 

\subsection{Validation with model Earth}

In order to validate that the neural network is able to retrieve
maps and rotation axes of planets that are not included in the training 
data, we used calculated data of the cloudy model planet Earth 
shown in Fig.~\ref{fig_earth_with_clouds}. 
The fluxes were computed using bidirectional reflection with 
polarization for a face-on orbit ($i=0^\circ$) with rotation axis 
coordinates $(0.918, 0.281, -0.281)$, such that the tilt is $23.4^\circ$.
Noise corresponding to $N_\text{max}= 10^4$ photons was added to 
the fluxes and they were normalized to a maximum value of 1. 
The network trained on planets with rotation axes with 
$y \geq 0$ predicts a rotation axis of $(0.904, 0.170, -0.331)$ 
for the light curves, corresponding to an MSE of $0.0050$, similar 
to the MSE obtained from the validation for face-on orbits.
We conclude that the neural network has not `memorized' 
the maps or rotation axes in the training data and 
constrains the rotation axis as intended.

%%%%%%%%%%%%%%%%%%%%%%%%%%%%%%%%%%%%%%%%%%%%%%%%%%%%%%%%%%%%%%%%%%%%%%%%%%%%%%
%--------------------------------------------------------------------------
% Figure 16
%--------------------------------------------------------------------------
\begin{figure}[htb!]
\centering
\includegraphics[width=.6\textwidth]{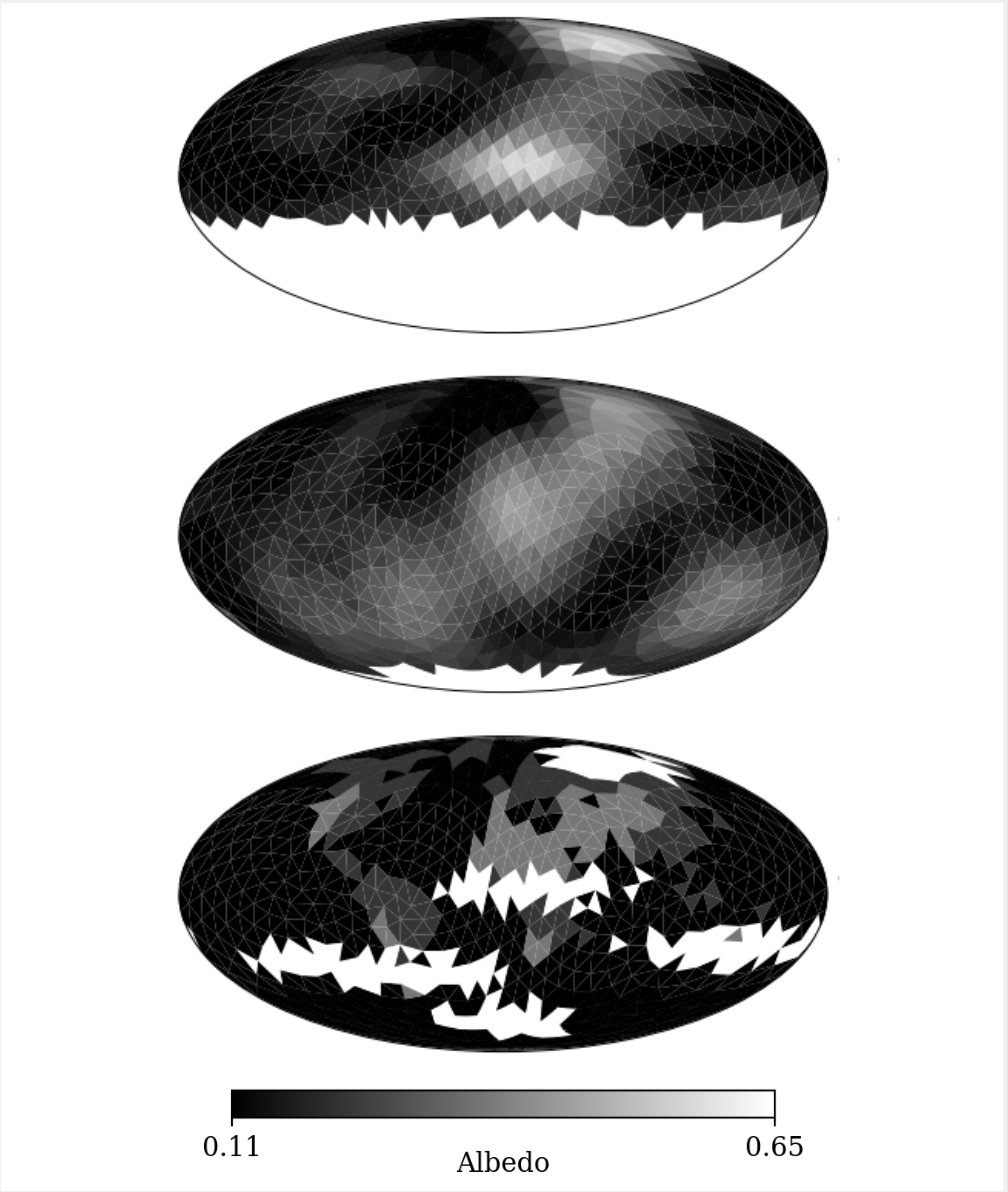}
\caption{Examples of albedo map retrievals for the best (top) and worst 
         (middle) geometries from Fig.~\ref{fig:axis_losses}
         using absolute light curves for $\lambda=550\ \text{nm}$.
         In these cases, $681$ and $917$ facets are visible, respectively.
         The map outside the visible region is white.
         The original map (bottom) is the model Earth 
         (at $\lambda=550\ \text{nm}$). 
         The MSE for these two retrievals is $0.0286$ and $0.0129$, respectively, 
         which is similar to the MSE of the validation data for these geometries
         (see Fig.~\ref{fig:axis_losses}).
    }
\label{fig:three_layer_retrievals}
\end{figure}
%--------------------------------------------------------------------------

%--------------------------------------------------------------------------
% Figure 17
%--------------------------------------------------------------------------
\begin{figure}[b!]
\centering
\includegraphics[width=0.6\textwidth]{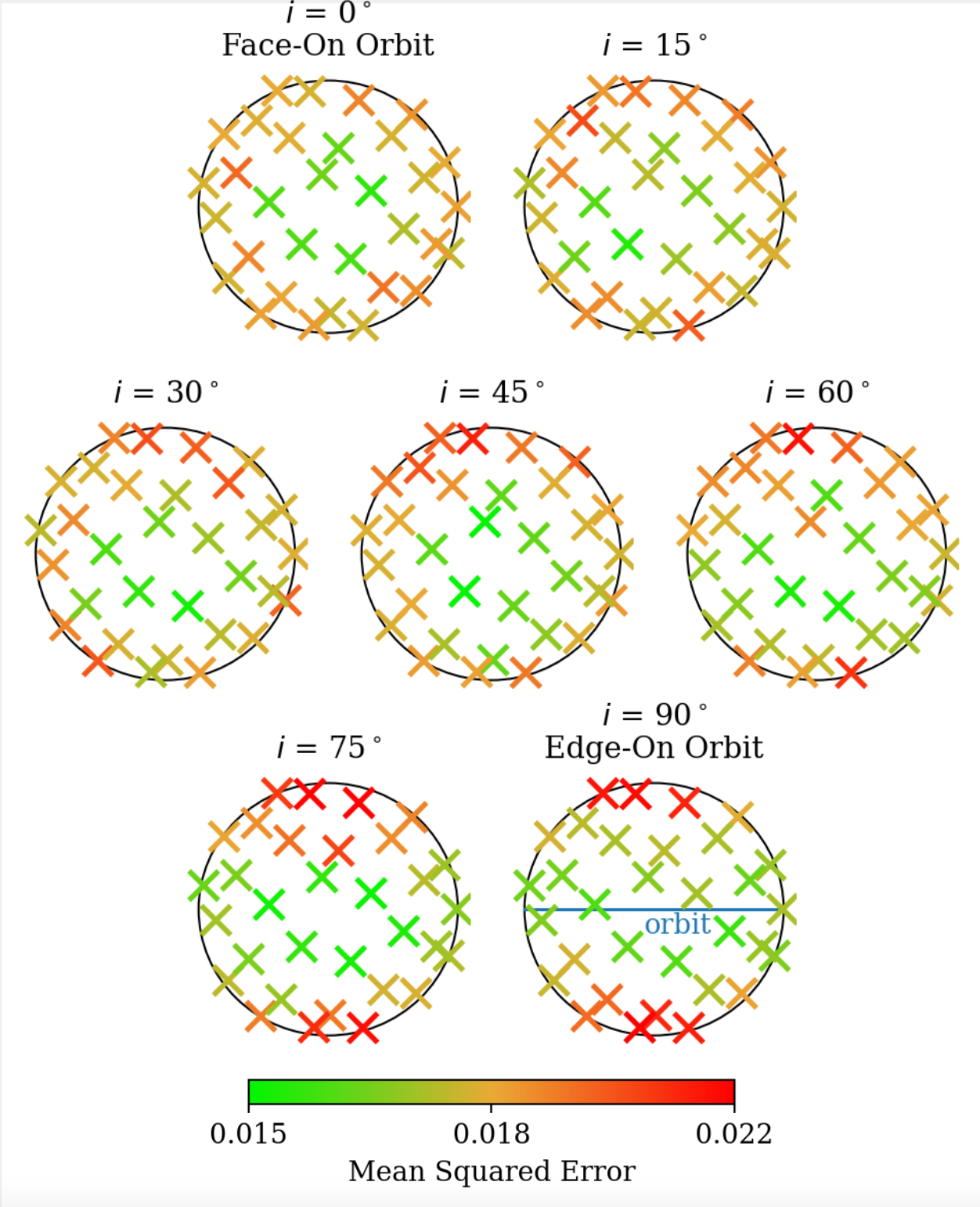}
\caption{Accuracy of albedo retrievals using the last 
         architecture from Table~\ref{tab:albedo_architectures} 
         for various combinations of rotation axis and inclination. 
         The axes are shown from the observer's perspective.
         Only the 32 axes with $x > 0$ (on the front-side) 
         are shown since the other 32
         axes are 
         mirrors with the rotation epochs in reverse order.
         The highest loss is found for $i = 90^\circ$ and an 
         axis of $(0.34, -0.11, -0.94)$, and the smallest loss 
         for $i = 60^\circ$ and the axis of $(0.93, -0.20, -0.30)$.}
\label{fig:axis_losses}
\end{figure}
%--------------------------------------------------------------------------

%--------------------------------------------------------------------------
% Table 3
%--------------------------------------------------------------------------
\begin{table}[htb!]
\centering
\caption{Tested architectures for retrieving facet albedos.}
\begin{tabular}{rcc} %{lll}
\hline\hline
Model Architecture   & Val.\ Loss & \# Epochs \\
\hline
$681$ Nodes & $0.0186$ & $413$ \\
\hline
\begin{tabular}[r]{@{}r@{}} $64$ Nodes \\
$681$ Nodes
\end{tabular} & $0.0164$ & $249$  \\
\hline
\begin{tabular}[r]{@{}r@{}}Periodic Convolution \\ $64$ Nodes \\ $681$ Nodes
\end{tabular} & $0.0146$ & $383$ \\
\hline
\end{tabular}
\tablefoot{These architectures are tested for the best combination of inclination and rotation axis (shown in Fig.~\ref{fig:axis_losses}), for which $681$ facets are visible. The inputs to the neural network are $64$ Lambertian fluxes ($8$ orbit locations $\times 8$ rotation phases) for the wavelength of $550\ \text{nm}$. The architectures are trained until the validation loss does not decrease for $10$ consecutive epochs. The periodic convolution uses a $1\times 3$ kernel, as shown in Fig.~\ref{fig:periodic_convolutions}.}
\label{tab:albedo_architectures}
\end{table}
%--------------------------------------------------------------------------

%--------------------------------------------------------------------------
% Table 4
%--------------------------------------------------------------------------
%\input{tab_albedo_val_losses}
\begin{table}[h]
\caption{Mean squared error (MSE) of the albedo retrievals.}
\centering
\begin{tabular}{rcc}
\hline\hline
$N_\text{max}$ & Model Earth MSE & Validation MSE  \\
\hline
$\infty$ & $0.0142$ & $0.0162$ \\
$10^5$ & $0.0146$ & $0.0185$ \\
$10^4$ & $0.0157$ & $0.0210$ \\
$10^3$ & $0.0184$ & $0.0240$ \\
$10^2$ & $0.0262$ & $0.0280$ \\
$10\hphantom{^2}$ & $0.0284$ & $0.0336$ \\
\hline
\end{tabular}
\label{tab:albedo_val_losses}
\tablefoot{The architecture from Fig.~\ref{fig:albedo_network} is trained on 
           normalized light curves to produce albedo maps and then fed 
           light curves of our model Earth and validation light curves 
           to find these errors. Compare with
           Fig.~\ref{fig:albedo_noise_retrievals} for visual indication 
           of different MSE levels.}
\end{table}
%--------------------------------------------------------------------------

%--------------------------------------------------------------------------
% Figure 18
%--------------------------------------------------------------------------
\begin{figure*}[t!]
\centering
\includegraphics[width = \textwidth]{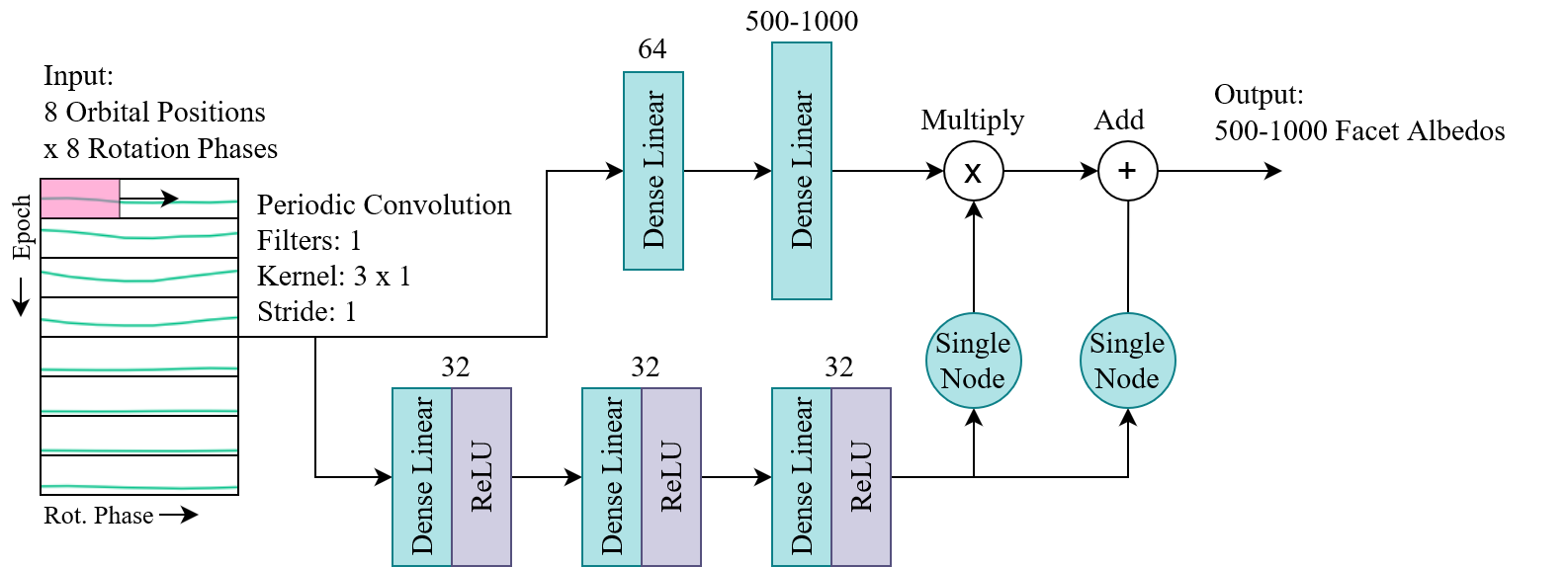}
\caption{The architecture we use to retrieve 
             planetary albedo maps from normalized phase curves.
             Several logic layers with ReLU activation
             functions are used to scale the output map by adding and
             multiplying by two single values. The scaling part of the
             network decreases the loss from $0.0197$ to $0.0162$ (for the
             ideal geometry as shown in Fig.~\ref{fig:axis_losses}). The
             output of the neural network is between $500$ and $1000$ 
             values, depending on the number of facets that is visible 
             for the specific geometry. Neurons in the bottom layers 
             have biases but neurons in the top layers do not.}
\label{fig:albedo_network}
\end{figure*}
%--------------------------------------------------------------------------
%--------------------------------------------------------------------------
% Figure 19
%--------------------------------------------------------------------------
\begin{figure}[h!]
\centering
\includegraphics[width=0.6\textwidth]{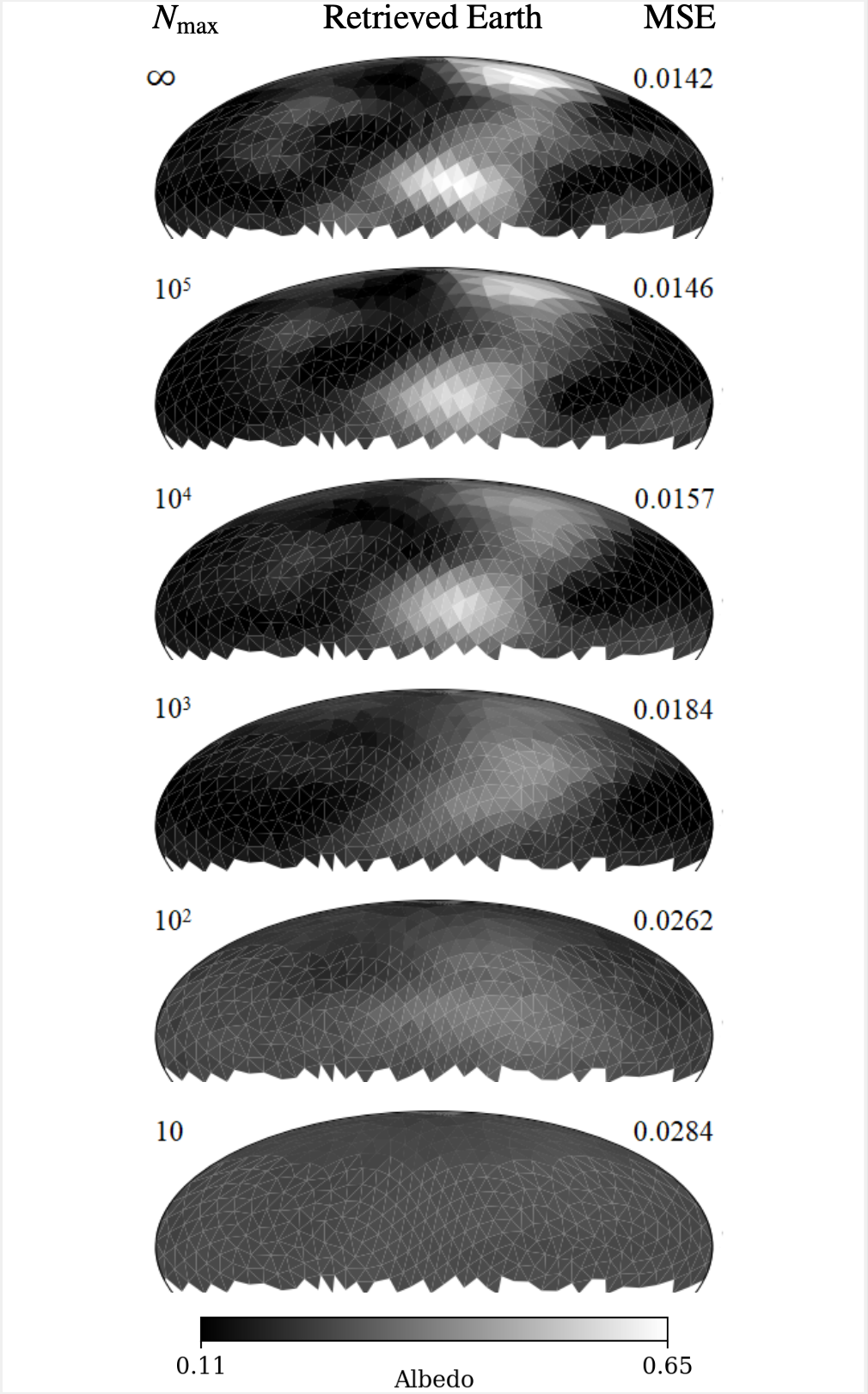}
\caption{Albedo retrievals of an Earth-like planet for different 
         noise levels ($N_{\rm max}=\infty$ is noise-free), using the optimal 
         configuration from Fig.~\ref{fig:axis_losses}. 
         The architecture from Fig.~\ref{fig:albedo_network} for total 
         flux curves computed using Lambertian reflecting model planets 
         is used. The actual Earth albedo map is shown in 
         Fig.~\ref{fig:three_layer_retrievals}.
         The MSE is computed by comparing the retrieved map
         with the map of the model Earth.
         }
\label{fig:albedo_noise_retrievals}
\end{figure}
%--------------------------------------------------------------------------

%%%%%%%%%%%%%%%%%%%%%%%%%%%%%%%%%%%%%%%%%%%%%%%%%%%%%%%%%%%%%%%%%%%%%%%%%%%
\section{Results: retrieval of albedo maps}
\label{sect_albedo_maps}

%%%%%%%%%%%%%%%%%%%%%%%%%%%%%%%%%%%%%%%%%%%%%%%%%%%%%%%%%%%%%%%%%%%%%%%%%%%
\subsection{Absolute light curves}
\label{sec:absolute_light_curves}

In this section, we attempt to replicate the results of other authors, retrieving planetary albedo maps based on absolute light curves computed using Lambertian reflection \citep[see for example][]{fujii_2012,sot_dynamic,2D_alien_map,exocartographer,nn_cartography}. Since information about facets that never face the observer is not present in the light curves, the neural network cannot retrieve the albedos of those facets. The number of output albedos values is thus between 500 (the axis is parallel to the line of sight and only half of the surface becomes visible) and 1000 (the axis is perpendicular to the line of the sight and the whole surface becomes visible). Note that every facet that becomes visible also becomes illuminated while being visible.

Since the problem is linear in nature, the first architecture that is tried is a single layer of nodes (see Table~\ref{tab:albedo_architectures}), meaning that each facet output is a linear combination of the $64$ flux values ($8$ orbit locations $\times~8$ rotation phases). Since no facet is biased towards a high or lower albedo, it is found that including biases in the nodes decreases the retrieval accuracy.

The neural network is trained using Lambertian, absolute flux curves for an orbital inclination of $60^\circ$ and a rotation axis of $(0.93, -0.20, -0.30)$ (this geometry is found to be best in Sect.~\ref{sec:axis_losses}). The wavelength chosen for the retrievals is $550~\text{nm}$, since at this wavelength it is easiest to distinguish the four surface types from each other (see Fig.~\ref{fig_surface_albedos}). Furthermore, $10\%$ of the data is used as validation data, the batch size of the training is set to $32$ (found by trial and error to be optimal), the network is trained until the loss does not decrease for $10$ consecutive epochs and the loss function used is the mean squared error (MSE) since this is a regression problem. These training parameters are used for all architectures discussed in this section.

The validation loss of the single layer after $413$ training epochs is $0.0186$. By trial and error, it is found that adding another layer of $64$ nodes decreases the MSE to $0.0164$ while also roughly halving the number of training epochs to $249$. Adding more dense layers than this does not further decrease the validation loss of the network. However, the accuracy can be improved by including a periodic convolution (described in Sect.~\ref{subsec:periodic_convolutions}) with a kernel size of $1\times 3$ before the two densely connected layers. The MSE of this model when applied to the validation data is equal to $0.0146$. Two example retrievals are shown in Fig.~\ref{fig:three_layer_retrievals} for a model planet Earth in the best and worst configurations of inclination and rotation axis shown in Fig.~\ref{fig:axis_losses}.

%%%%%%%%%%%%%%%%%%%%%%%%%%%%%%%%%%%%%%%%%%%%%%%%%%%%%%%%%%%%%%%%%%%%%%%%%%%
\subsection{Retrieval accuracies for different geometries}
\label{sec:axis_losses}

In this section, the effect of the orbital inclination angle 
and the rotation axis orientation on the albedo map retrievals 
is investigated. Only half of the possible $64$ rotation axes is 
discussed since all axes with $x \leq 0$ can be reflected in the 
$yz$ plane to create a new axis with the same observations in reverse
order, resulting in the same retrieval accuracy 
(see Fig.~\ref{fig_orbital_positions} for the definition of the
$(x,y,z)$ coordinate system).

To this end, the final model from Table~\ref{tab:albedo_architectures} 
is re-trained for all $32$ axes with $x \geq 0$ for each of the seven 
inclinations ($0^\circ$, $15^\circ$, $30^\circ$, $45^\circ$, $60^\circ$, 
$75^\circ$ and $90^\circ$). The resulting MSEs of the validation data 
are plotted in Fig.~\ref{fig:axis_losses}. Note that the number of facets 
that are retrieved varies for each rotation axis, since not all facets 
become visible.

We find that a rotation axis angle (angle between the rotation axis and orbital normal vector) near $90^\circ$ provides the best retrieval accuracy for edge-on and near-edge-on orbits (i.e. $i = 75^\circ$ and $i = 90^\circ$). For edge-on orbits (in the $xy$ plane), the orbital movement modulates the signal across the $y$ (vertical) axis of the planet's surface. When the rotation axis has a large component in the orbital plane, this provides modulation in the opposite ($z$) direction. Conversely, when the axis is normal to the orbital plane the modulation due to the rotation of the planet is in the same direction as the modulation due to the orbital movement. edge-on orbits with rotation axes normal to orbital plane thus have the worst accuracy of all combinations. The worst geometry that is studied is for an edge-on orbit and a rotation axis of $(0.34, -0.11, -0.94)$, with an MSE of $0.022$.

For face-on and near-face-on orbits, the orbital movement of the planet modulates the signal in both the $y$ and $z$ directions. The best retrieval accuracies are then found for axes near %to
the normal of the orbital plane that also modulate in both directions. Axial tilts near to $90^\circ$ modulate in only one direction and thus show slightly worse results. The best geometry is found for an inclination of $60^\circ$ and an axis of $(0.93, -0.20, -0.30)$, as in this case the orbit and rotation axis both modulate across two perpendicular directions. The rotation axis angle is $78^\circ$, so the rotation axis lies close to the orbital plane, roughly in the direction of the observer. The MSE for this case is $0.015$ and this geometry is chosen for many of the remaining example retrievals in this paper.

%%%%%%%%%%%%%%%%%%%%%%%%%%%%%%%%%%%%%%%%%%%%%%%%%%%%%%%%%%%%%%%%%%%%%%%%%%%
\subsection{Relative (normalized) light curves}
\label{sec:relative_curves}

Since the radius of an exoplanet is difficult to constrain from 
direct detections (a small, bright planet can have the same flux
as a large, dark one), we use flux curves that are normalized to 1.0
for our retrieval algorithms for directly observed exoplanets. 
To this end, we use a modified architecture as shown in 
Fig.~\ref{fig:albedo_network} in which the outputs of the periodic
convolution are used to estimate the overall brightness of the 
albedo maps by three densely connected layers with $32$ nodes each. 
Two single nodes (which each output a single value) are then 
multiplied and added to the albedo map to scale the map to the 
estimated brightness. The densely connected layers have ReLU 
activation functions, which are a special case of PReLU 
(Eq.~(\ref{eq:prelu}) with $a=0$). 
Based on our experience with this architecture, 
PReLU layers where $a$ is 
actively trained instead of ReLU layers do not provide better 
results and increase the number of epochs needed to reach a
similar validation loss.

When using the network designed for absolute 
flux curves (Table~\ref{tab:albedo_architectures}) and training it 
with normalized flux curves, the MSE increases from $0.0146$ to 
$0.0197$. Using the scaling discussed above decreases the MSE 
from $0.0197$ to $0.0162$. 
This shows that the network can effectively estimate the 
map's brightness.

%%%%%%%%%%%%%%%%%%%%%%%%%%%%%%%%%%%%%%%%%%%%%%%%%%%%%%%%%%%%%%%%%%%%%%%%%%%
\subsection{Effects of noise}
\label{sec:effects_of_noise_albedos}

In this section, we study the effect of noise on the albedo map 
retrievals. We train the architecture from Fig.~\ref{fig:albedo_network} 
with normalized, Lambertian light curves with different levels of 
photon noise added (see Sect.~\ref{sec:noise}). 
The MSE of the validation data and the retrieved model Earth maps are 
shown in Table~\ref{tab:albedo_val_losses} and in 
Fig.~\ref{fig:albedo_noise_retrievals}. 
For all noise levels the retrieval of the model Earth is 
significantly better than that of the validation data. 
This could be because the model Earth is largely covered by 
vegetation and ocean, which have similar albedo's for the 
studied wavelength of $550\ \text{nm}$ when compared with a 
combination of clouds and ocean, for example.
Homogeneous planets may be easier to retrieve since there are less 
hard-to-predict details with large penalties for the loss function.

For the case without noise ($N_\text{max}=\infty$), the clouds, 
the Sahara and North America can be clearly distinguished. These 
features disappear at a noise level corresponding to $N_\text{max} \leq 10^3$ photons. The results shown here are comparable to results from 
\citet{fujii_2012,exocartographer,2D_alien_map,sot_dynamic,nn_cartography}, 
although a precise comparison is difficult due to differences 
in geometry, noise models, number of observations and the specific map.

%%%%%%%%%%%%%%%%%%%%%%%%%%%%%%%%%%%%%%%%%%%%%%%%%%%%%%%%%%%%%%%%%%%%%%%%%%%

%%%%%%%%%%%%%%%%%%%%%%%%%%%%%%%%%%%%%%%%%%%%%%%%%%%%%%%%%%%%%%%%%%%%%%%%%%%%%%
%--------------------------------------------------------------------------
% Figure 21
%--------------------------------------------------------------------------
\begin{figure*}[t!]
\centering
\includegraphics[width=\textwidth]{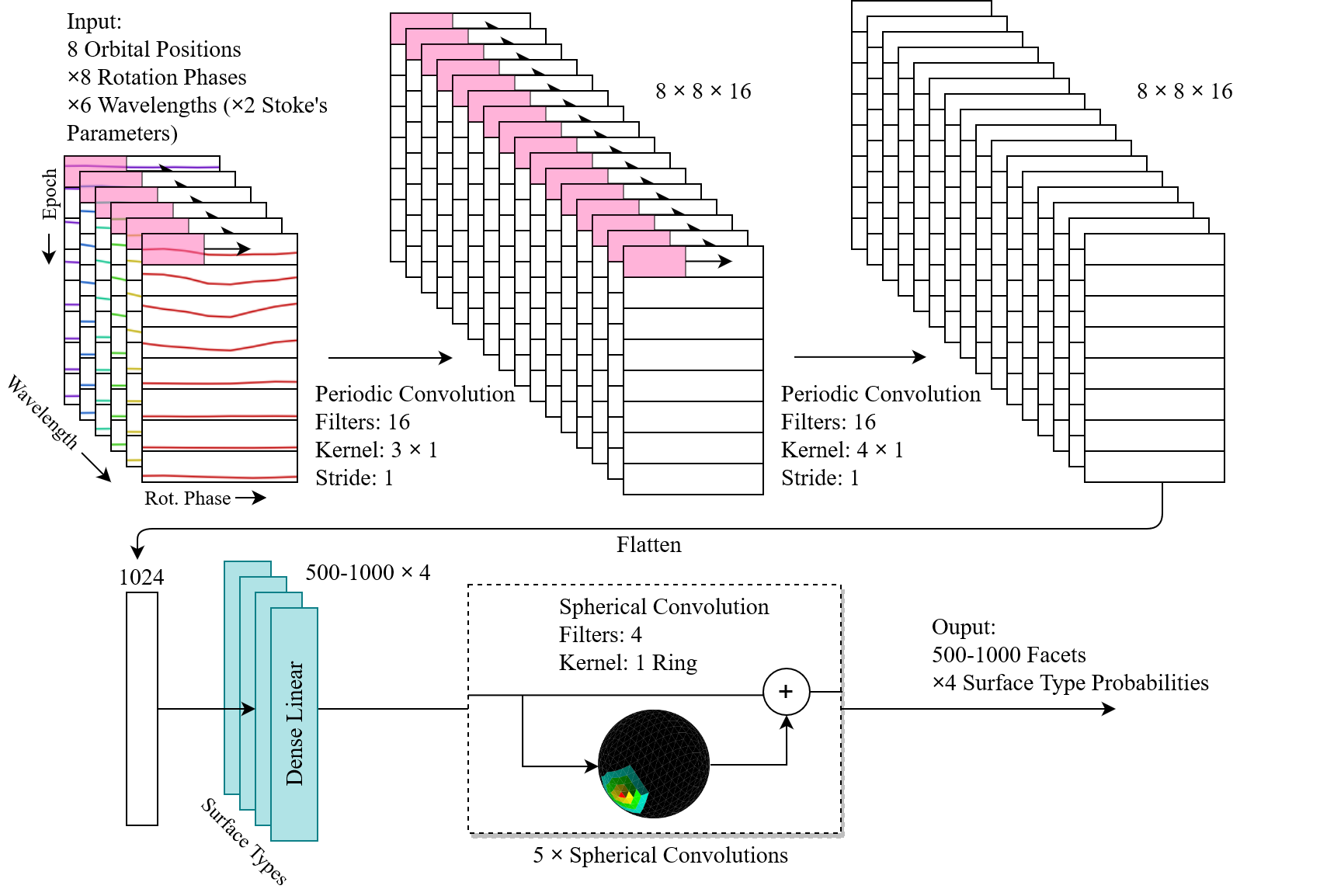}
\caption{The architecture for classifying facets on a planet as one of 
         four surface types. Several periodic convolutions recognize patterns 
         in the curves similar to the rotation axis retrieval network 
         (Fig.~\ref{fig:final_nn}). To maintain a high level of resolution 
         of rotation phases, no down-sampling is used in the convolutions. 
         Four dense layers with $500$ to $1000$ nodes estimate the probability 
         of each surface type for all visible facets. We apply five 
         spherical convolutions in series and add the outputs to achieve 
         the final result. We used four filters in the spherical convolutions 
         to match the dimensions.}
    \label{fig:classification_network}
\end{figure*}
%--------------------------------------------------------------------------

%--------------------------------------------------------------------------
% Figure 22
%--------------------------------------------------------------------------
\begin{figure*}[ht]
\centering
\includegraphics[width=\textwidth]{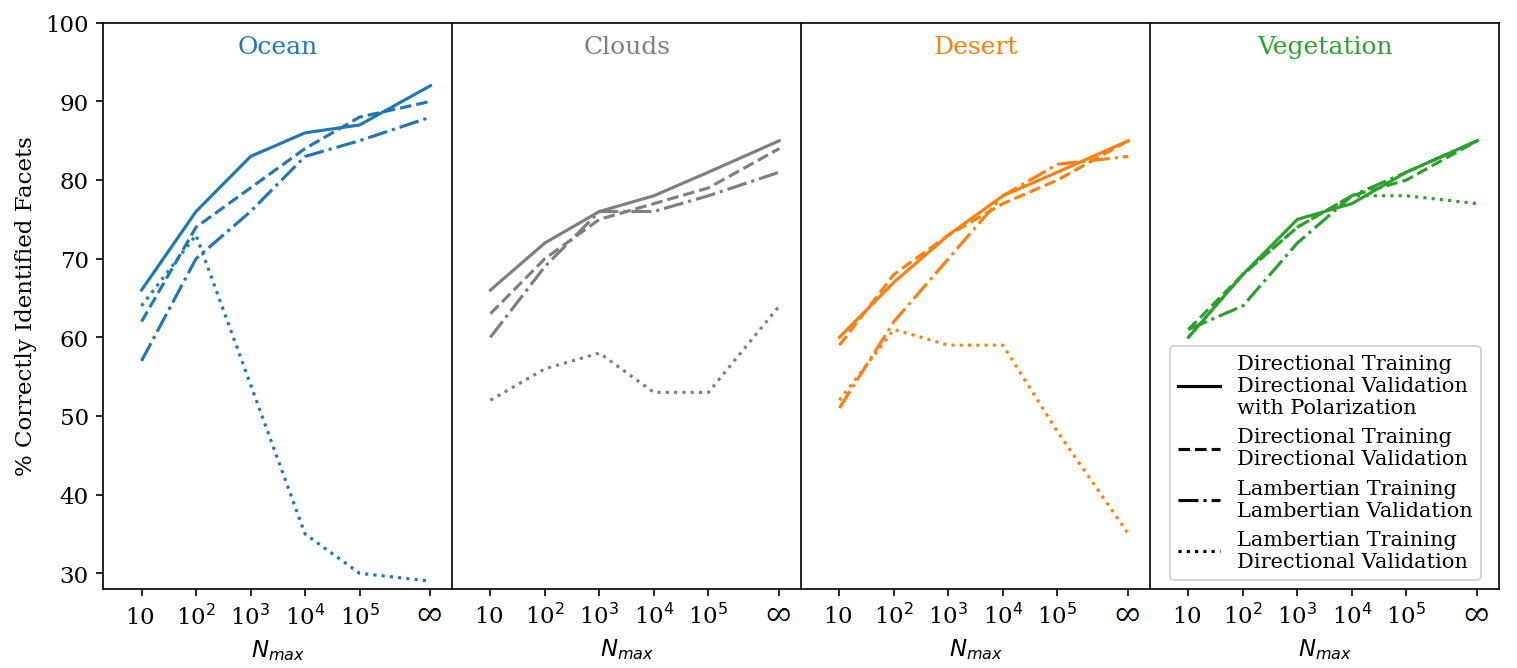}
\caption{The percentages of correctly retrieved facets in the validation 
         data for different noise levels and different combinations of 
         training data (Lambertian or bidirectional reflection) 
         and validation data (Lambertian or birdirectional reflection).
         Ocean, clouds, desert and 
         vegetation facets show maximal percentages of $92\%$, $85\%$, 
         $85\%$, and $85\%$, respectively. 
         The architecture in Fig.~\ref{fig:classification_network} was 
         trained for the best combination of orbital inclination and rotation 
         axis shown in Fig.~\ref{fig:axis_losses}.}
\label{fig:classification_accuracies}
\end{figure*}
%--------------------------------------------------------------------------

%--------------------------------------------------------------------------
% Figure 23
%--------------------------------------------------------------------------
\begin{figure*}[hb!]
\centering
\includegraphics[width=\textwidth]{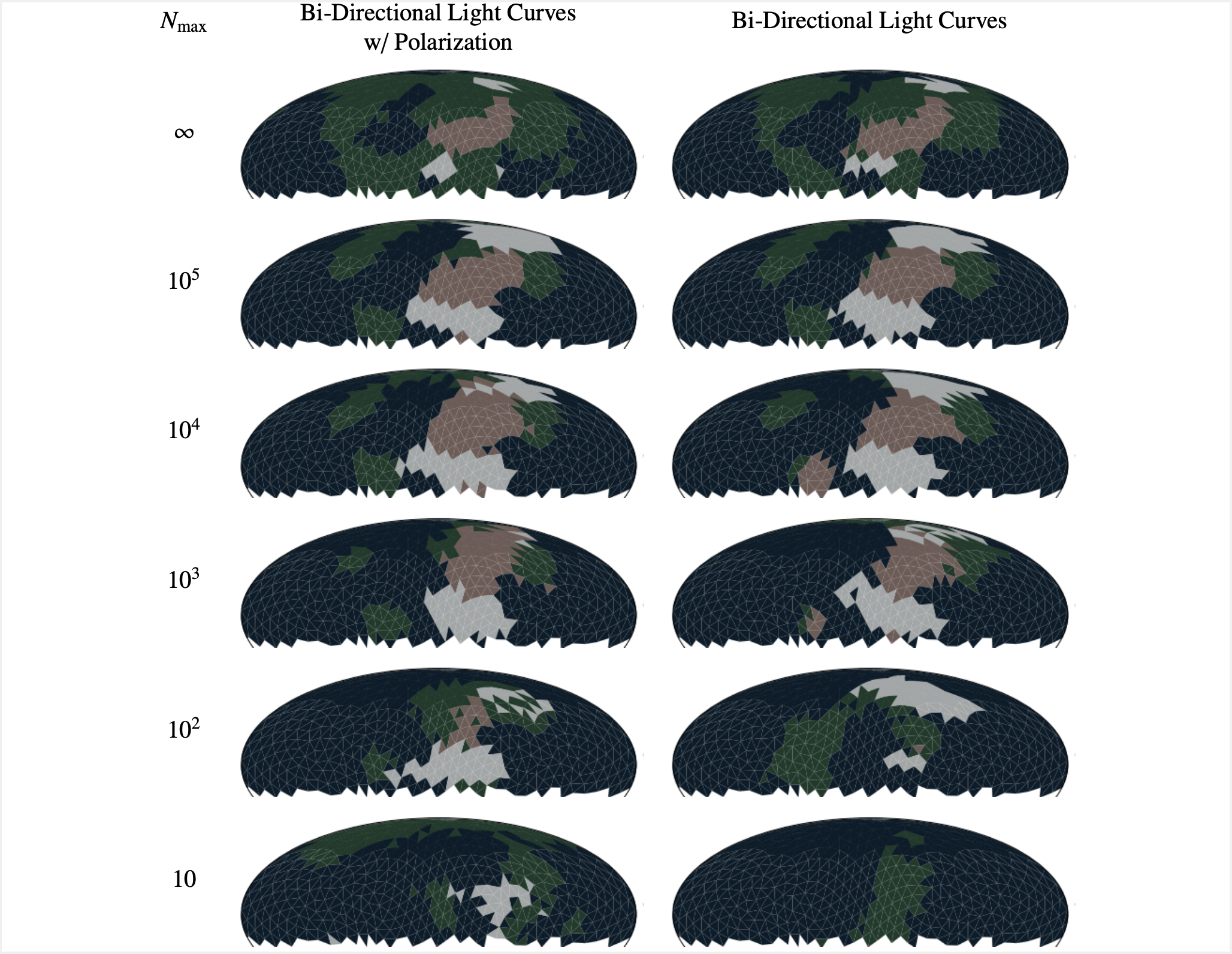}
\caption{Retrieved maps of the model Earth (bottom of 
         Fig.~\ref{fig_earth_with_clouds}) for different levels of 
         photon noise, using directional light curves with (left) and 
         without polarization (right).
         The architecture from Fig.~\ref{fig:classification_network} was used
         and trained for the optimal configuration of Fig.~\ref{fig:axis_losses}.
         For $N_\text{max}= \infty$ photons, there is no noise.}
\label{fig:retrieved_earth}
\end{figure*}
%--------------------------------------------------------------------------

%--------------------------------------------------------------------------
% Figure 20
%--------------------------------------------------------------------------
\begin{figure}[t!]
\centering
\includegraphics[width = 0.5\textwidth]{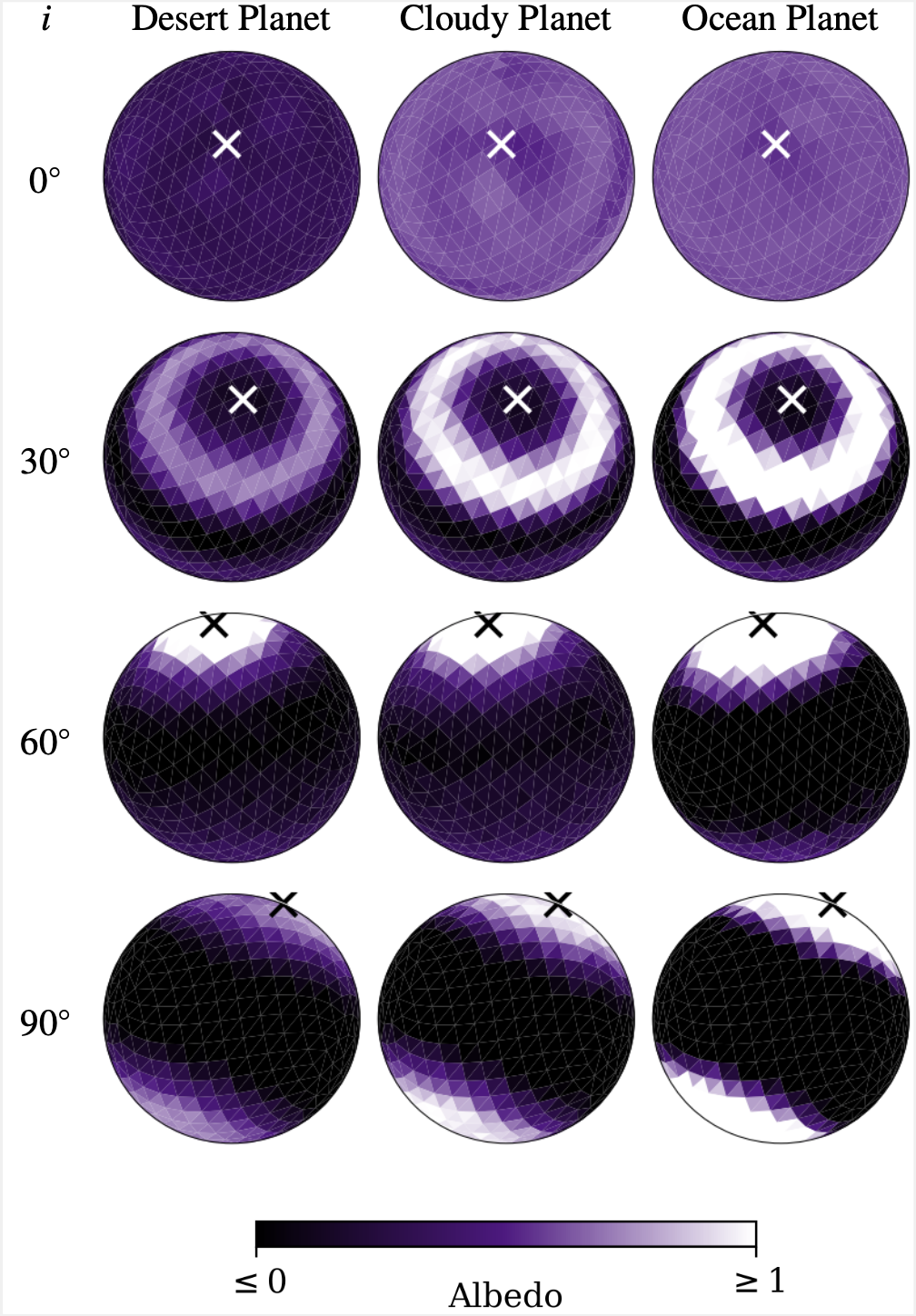}
\caption{Albedo-maps of bidirectionally reflecting planets at 
         $\lambda=400$~nm for homogeneous desert, cloud, and ocean planets
         retrieved by the network architecture of Fig.~\ref{fig:albedo_network} 
         that was trained on Lambertian reflecting planets, for different
         orbital inclination angles $i$.
         The rotation axis for each planet (indicated with a white or
         black cross) was chosen from the set shown in 
         Fig.~\ref{fig:axis_losses} to minimize the difference to the 
         normal of the orbital plane. 
         The tilt angles of the axes are $15^\circ$, $5^\circ$, $11^\circ$ and 
         $23^\circ$ in order of ascending inclination.}
\label{fig:albedo_network_directional_curves}
\end{figure}
%--------------------------------------------------------------------------

%--------------------------------------------------------------------------
% Figure 24
%--------------------------------------------------------------------------
\begin{figure}[hb]
\centering
\includegraphics[width = 0.6\textwidth]{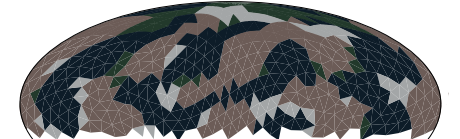}
\caption{The map of the model Earth as retrieved from bidirectional
         light curves by the network that was trained with Lambertian 
         reflecting curves. The orbital inclination and rotation axis of 
         the planet are the optimal configuration shown in  Fig.~\ref{fig:axis_losses}.
         The actual map is shown in Fig.~\ref{fig_earth_with_clouds}.}
\label{fig:lambertian_model_directional_curves}
\end{figure}
%--------------------------------------------------------------------------

%--------------------------------------------------------------------
% Figure 25
%--------------------------------------------------------------------
\begin{figure}[h!]
\centering
\includegraphics[width=87mm]{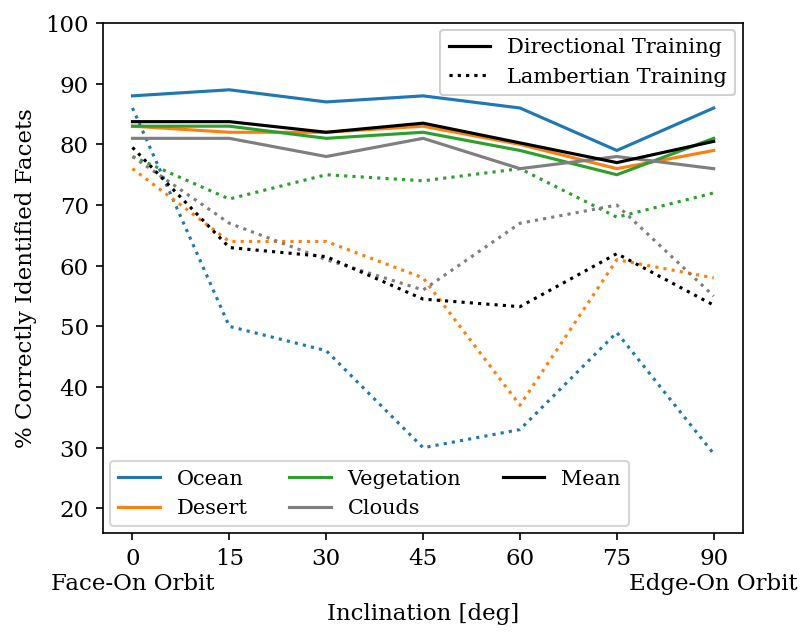}
\caption{The accuracy of the retrieval algorithms that trained on 
         Lambertian (dashed) and bidirectionally (solid) reflecting 
         planets, when applied to light curves of bidirectionally reflecting
         planets, as functions of the orbital inclination angle $i$. 
         The rotation axes for each inclination angle are chosen to 
         minimize the difference to the normal on the orbital plane (see 
         Fig.~\ref{fig:albedo_network_directional_curves} for examples).}
\label{fig:inclination_classification_accuracies}
\end{figure}
%--------------------------------------------------------------------

%%%%%%%%%%%%%%%%%%%%%%%%%%%%%%%%%%%%%%%%%%%%%%%%%%%%%%%%%%%%%%%%%%%%%%%%%%%
\section{Results: retrieval of surface-type maps}
\label{sect_surface_type_maps}

Since albedo maps do not fully describe the properties of a planet and 
do not identify unique non-Lambertian reflecting patterns due to 
oceans and clouds, we here retrieve for each facet that covers the planet,
a surface type instead of an albedo, as in Sect.~\ref{sect_albedo_maps}. 
The four possible surface types for each facet are: 
ocean, vegetation, sandy desert, and clouds (while strictly speaking
clouds are not on the surface, the network assumes them to be).
For each facet, our neural network predicts a 
probability for each surface type.
When creating maps or checking the classification accuracy,  
each facet is assigned the surface type with the highest probability.

%%%%%%%%%%%%%%%%%%%%%%%%%%%%%%%%%%%%%%%%%%%%%%%%%%%%%%%%%%%%%%%%%%%%%%%%%%%
\subsection{Architecture}

To create surface type maps, the output dimensions of the 
neural network should equal the number of visible facets
(between $500$ and $1000$) times the number of possible surface 
types ($4$). To recognize patterns in the light curves, 
we use a similar periodic convolution as shown in Fig.~\ref{fig:final_nn}. 
Two periodic convolutions (see Sect.~\ref{subsec:periodic_convolutions})
with $16$ kernels of size $1\times 3$ and $1\times 4$ are applied to the
light curves. No down-sampling is done as the resolution along the 
rotation phase should be high to create accurate maps; the stride 
is thus 1. This means that we use a vector of size $1024$ after 
flattening the output of the periodic convolutions. 
This is followed by a dense layer for each surface type. 
Each dense layer has as many nodes as there are output facets.

Finally, we use spherical convolutions to regularize the surface map. 
With a ResNet approach \citep{resnet} the dense layer output is fed 
into the convolutions (four filters, one ring) and the output 
is added to the previous output. This is repeated 5 times to achieve the 
final output of the neural network. 
Before training, the spherical convolution kernels are initialized with 
zeros, to avoid interference with the learning of earlier layers.

%%%%%%%%%%%%%%%%%%%%%%%%%%%%%%%%%%%%%%%%%%%%%%%%%%%%%%%%%%%%%%%%%%%%%%%%%%%
\subsection{Training the surface-type network}

The neural network is trained for one combination of inclination 
and rotation axis at a time, so the training data consists of roughly 
$4\small{,}000\small{,}000 / 7 / 64 \approx 9\small{,}000$ light curves, 
of which $10\%$ are set aside to use for validation. 
The neural network is trained using the Adam optimization algorithm 
\citep{adam}, until the validation loss does not decrease for five %5
consecutive epochs. We find that a batch size (number of training curves 
fed into the neural network simultaneously) of $32$ works best, since 
larger batch sizes can cause the network to overshoot the local minima. 
This could be due to the gradient magnitudes in combination with the 
standard learning rate in Keras. 
We also find that training the neural network using the MSE
as the loss function gives the highest retrieval accuracies when
compared to the categorical cross entropy that is usually used 
for classification problems \citep{categorical_cross_entropy}. 
This might be due to the ill-defined nature of the exocartography 
inversion problem or to a sub-optimal network architecture.

%%%%%%%%%%%%%%%%%%%%%%%%%%%%%%%%%%%%%%%%%%%%%%%%%%%%%%%%%%%%%%%%%%%%%%%%%%%
\subsection{Retrieval accuracy}
\label{sec:map_retrieval_accuracies}

In this section, the retrieval accuracies of the neural network 
architecture shown in Fig.~\ref{fig:classification_network} are 
investigated. The retrieval accuracy of each surface type is plotted 
as a function of the noise level in Fig.~\ref{fig:classification_accuracies}.
The inclination and rotation axis chosen were the best combination 
as shown in Fig.~\ref{fig:axis_losses}. Since the inclination angle 
is $60^\circ$, the rainbow feature at $\alpha = 38^\circ$ is 
visible at two of the eight orbital positions and the ocean glint 
is also prominent at the two orbital positions for which 
$\alpha = 142^\circ$.

The network performs poorly when trained on Lambertian light curves 
while applied to directional light curves. 
This will be further discussed in
Sect.~\ref{sec:effect_of_lambertian_assumption}. 
The surface type that is correctly classified most often is ocean, 
due to its unique glint feature as well as its characteristic darkness. 
We verified that the high accuracy is (at least) partially due to the 
ocean glint since the Lambertian trained network applied to Lambertian light 
curves does not perform as well as the directional trained network applied to 
directional light curves. 
A similar effect is also seen for cloudy facets, which also have unique,
bidirectional reflection (see Fig~\ref{fig_phase_curves}). 
The accuracy of the classification of vegetation and desert surface types, 
which are modelled as Lambertian reflectors, does not increase when 
bidirectional curves instead of Lambertian curves are used. 
We find that excluding the light curves with a wavelength of 550~nm 
(the green bump) does not impact the retrieval accuracy of the surface 
maps in the absence of noise.

The retrieved maps of the Earth-like planet are shown in
Fig.~\ref{fig:retrieved_earth} for different noise levels. 
These maps have more detail than the albedo maps shown in 
Fig.~\ref{fig:albedo_noise_retrievals}, which validates our approach 
of retrieving surface types rather than albedos. 
In the absence of noise ($N_{\rm max}=\infty$), 
all continents can clearly be distinguished. 
At a noise level for photon numbers $N_\text{max} \leq 10^3$, 
the Americas disappear. 
Lastly, for $N_\text{max}= 10$, only the main surface types 
(vegetation and ocean) are retrieved. At this high noise level,
the retrieved map depends strongly on probabilistic noise contributions.

%%%%%%%%%%%%%%%%%%%%%%%%%%%%%%%%%%%%%%%%%%%%%%%%%%%%%%%%%%%%%%%%%%%%%%%%%%%

\section{Discussion}
\label{sect_discussion}

Our method of retrieving exoplanet surface maps is based on the assumption 
that the planet has an atmosphere with clouds and three surface types: 
ocean, desert and Earth-like vegetation. These are modelled with bidirectional
reflection for clouds and a rough ocean with glint. The planet has a fast 
rotation compared to the orbital period, as its surface is static. Here, 
we first discuss the results of using neural networks, then the results 
from our inclusion of bidirectional reflection and polarization in the signal, 
and list possible steps for an observational campaign and recommendations 
for future work.

\subsection{Optimizing the neural network architectures}

The rotation axis retrieval is non-linear and therefore a neural network with
non-linear activation functions was found to be most accurate. On the other 
hand, the neural network for creating surface-type maps was found to work best 
when only linear layers were used, contrary to our initial assumptions. The 
linear neural network is equivalent to a least squares method (with regularization)
which defines the ideal solution for a data set. It remains to be seen whether 
neural networks have advantages over least-squares solutions for the map 
retrieval with a constrained rotation axis.

Our model assumes a static map, and we thus do not include
seasonal variations, such as dynamical ice caps, hurricane clouds, and 
volcanic eruptions. Although non-static maps can be retrieved with a 
least-squares method, this is at the cost of spatial resolution 
\citep{sot_dynamic}.
Neural networks could be trained to make connections between planet 
characteristics, such as a relation between an obliquity of the planet's 
rotation axis and the varying size of a polar ice cap, or between cloud 
cover and oceans. They thus appear to be suitable for the retrieval of 
dynamic maps, and we expect them to prove their usefulness in future 
exocartography.

\subsection{Consequences of the Lambertian assumption}
\label{sec:effect_of_lambertian_assumption}

Since all other authors, to our knowledge, use Lambertian reflection for 
their retrievals, it is interesting to evaluate the validity of this 
Lambertian assumption for training a network when the actual 
data will be of planets that reflect non-Lambertian, i.e.\
that reflect bidirectionally. When retrieving the rotation axis, 
the network trained with Lambertian curves and applied to Lambertian 
reflecting model planets performs best of all 
combinations in Fig.~\ref{fig:incl_losses}:
the MSE is $0.0074$ for face-on orbits.
However, when this network that was trained with Lambertian model
planets is applied to the more realistic bidirectional light curves, 
the losses increase by roughly one order of magnitude for large inclination
angles. 
Thus evaluating the network's performance while using only 
Lambertian light curves leads to a false confidence in the retrieval 
algorithm's accuracy. 
When such a retrieval algorithm is used to retrieve the rotation axes of
bidirectionally reflecting planets in a face-on orbit, however, the 
MSE is $0.0117$, only $18\%$ larger than the MSE of $0.0099$ for the 
network that was trained on bidirectional curves. This confirms that 
in an edge-on orbit, where the phase angle is constant along the orbit,
the reflection appears to be mostly Lambertian because of the lack
of angular features.

The MSE of the retrievals by the networks trained with bidirectional 
models increases with increasing inclination angle.
Including polarization in the retrievals overall increases the 
retrieval accuracy. 
The beneficial effects of including polarization are 
smallest for $i=0^\circ$ and $15^\circ$. This can be explained 
since the reflected flux for these small inclinations appears to
be mostly Lambertian. The largest decrease in MSE when including 
polarization occurs for $i=60^\circ$, where including polarization 
decreases the loss by $26\%$.

To test the Lambertian assumption for albedo map retrievals, we use 
the architecture from Fig.~\ref{fig:albedo_network}, which provides 
comparable results to those 
by other authors by training on Lambertian light curves and testing on
directional light curves (i.e.\ light curves computed using directional 
reflection). As can be seen in Fig.~\ref{fig:albedo_network_directional_curves},
strong concentric artefacts about the pole appear for all inclinations besides 
for a face-on orbit. Each of the planets in the figure is a homogeneous planet 
and should be retrieved as such. Since the planets have the same flux at all
rotation phases, the artefacts are purely due to the overestimation of the 
flux at low phase angles and underestimation of the flux at high phase 
angles when using the Lambertian assumption (see Fig~\ref{fig_phase_curves}). 
Since most exoplanets are not in a face-on orbit, these errors demonstrate a 
need for new retrieval algorithms that use directional light curves to map 
planet surfaces.

For surface-type map retrievals, the Lambertian trained neural network 
yields large errors when applied to 
bidirectional light curves, as can be seen in Fig.~\ref{fig:classification_accuracies}. As an example,
Fig.~\ref{fig:lambertian_model_directional_curves} shows the map as 
retrieved from bidirectional light curves of the model Earth 
(see Fig.~\ref{fig_earth_with_clouds}) by a neural network that was
trained on Lambertian light curves. 
Amongst others, the network predicts a large fraction of 
desert facets where there should be dark ocean facets.

Since so far we have only studied the ideal geometry from 
Fig.~\ref{fig:axis_losses}, we also used other inclinations to 
investigate if that improves the retrieval of the Lambertian trained 
network, like with the rotation axis and albedo map retrievals for 
a face-on orbit. 
Figure~\ref{fig:inclination_classification_accuracies} shows the
retrieval accuracy for Lambertian and bidirectional trained networks
as functions of the inclination angle, using a rotation axes tilt close 
to $0^\circ$. 
As seen in Fig.~\ref{fig:inclination_classification_accuracies}, 
the Lambertian trained
network is only reliable for orbits that are face-on or close to face-on:
only for $i \leq 10^\circ$ all surface types have a classification 
accuracy larger than $60\%$, and even then, the accuracy of the 
directional network is considerably higher and thus preferred.

The only surface type for which the classification accuracy of 
the Lambertian trained network is higher than $65\%$ for all 
inclination angles is vegetation. 
Indeed, the Lambertian network can use the red edge feature in the albedo 
(see Fig.~\ref{fig_surface_albedos}) to recognize vegetation. 
Since the atmospheric influence on the signal is very small at the
large wavelengths of the red edge, and because vegetation is modelled 
as a Lambertian reflector, vegetation facets appear very similar 
in directional and Lambertian light curves 
(see Fig~\ref{fig_phase_curves}).

%%%%%%%%%%%%%%%%%%%%%%%%%%%%%%%%%%%%%%%%%%%%%%%%%%%%%%%%%%%%%%%%%%%%%%%%%%%

\subsection{Benefits of polarization}
\label{sec:benefits_of_polarization}
% This was 6.5:
The effects of including polarization on the surface-type map retrievals can be seen 
in Table~\ref{tab:confusion_matrix_no_pol} which shows the 
confusion matrices for the bidirectional neural networks 
with and without polarization.
Confusion matrices visualize the performance of classification 
schemes by showing for each actual class the fractions of the
possible classes that are retrieved, thus, the fraction of
e.g.\ ocean facets that is retrieved as ocean, desert, vegetation,
and cloud facets.

A comparison of the confusion matrices shows that the retrieval 
accuracy of the desert and vegetation-covered surface facets is 
insensitive to including the polarization signal to the model 
observations and the retrieval algorithm. 
This is to be expected since both surface types are 
modelled as Lambertian, non-polarizing surfaces. 
Indeed, the retrieval accuracy of ocean and cloud facets is 
increased by 2\% and 1\%, respectively, due to the unique polarization 
signatures of the ocean glint and the rainbow. 
Including polarimetry on a future exoplanet characterisation 
telescope would thus increase the ability to map oceans and
clouds on exoplanets.

%--------------------------------------------------------------------------
% Table 5
%--------------------------------------------------------------------------
%\input{tab_confusion_matrix_no_pol.tex}
\newcolumntype{A}{>{\centering\arraybackslash} p{1cm} }
\definecolor{mygreen}{RGB}{0,129,64} 
\definecolor{myred}{RGB}{129,0,0}

\begin{table}[b!]
\caption{Confusion matrices for the neural network shown in 
 Fig.~\ref{fig:classification_network}.}
\centering
\label{tab:confusion_matrix_no_pol}
Without polarization
\begin{tabular}{l|AAAA}
\hline\hline
\setlength\tabcolsep{0pt}
\diagbox{Pred.}{Actual} & Ocean & Desert & Veget. & Clouds \\
\hline
Ocean  & ${0.90}$ & $0.06$ & $0.03$ & $0.05$ \\
Desert & $0.05$ & ${0.85}$ & $0.08$ & $0.06$ \\ 
Veget.\ & $0.02$ & $0.05$ & ${0.85}$ &  $0.05$ \\ 
Clouds & $0.03$ & $0.04$ & $0.04$  & ${0.84}$ \\
\hline
\end{tabular}
\\
\mbox{} \vspace{5mm} \mbox{}
\\
With polarization
\begin{tabular}{l|AAAA}
\hline\hline
\setlength\tabcolsep{0pt}
\diagbox{Pred.}{Actual} & Ocean & Desert & Veget. & Clouds \\
\hline
Ocean & \textcolor{mygreen}{${0.92}$} & $0.06$ & $0.03$ & $0.05$ \\
Desert & \textcolor{myred}{$0.04$} & ${0.85}$ & $0.08$ & \textcolor{myred}{$0.05$}   \\ 
Veget. & \textcolor{myred}{$0.01$} & $0.05$ & ${0.85}$ & \textcolor{myred}{$0.04$} \\ 
Clouds & $0.03$ & \textcolor{myred}{$0.03$} & $0.04$ & \textcolor{mygreen}{${0.85}$} \\
\hline
\end{tabular}
\tablefoot{The accuracy of the retrieval using directional light 
           curves without polarization (top), and with polarization 
           (bottom), for planets with the best configuration from
           Fig.~\ref{fig:axis_losses},
           without noise. Shown are the fractions of each surface types' 
           facets (columns) that are classified as a specific type (rows).
           The diagonals show the correct classification of each 
           surface type (cf.\ Fig.~\ref{fig:classification_accuracies}
           for $N_\text{max} = \infty$).
           An increase or decrease with respect to the matrix without
           polarization (top) is marked in green and red in the matrix with
           polarization (bottom), respectively.} 
\end{table}
%--------------------------------------------------------------------------

%%%%%%%%%%%%%%%%%%%%%%%%%%%%%%%%%%%%%%%%%%%%%%%%%%%%%%%%%%%%%%%%%%%%%%%%%%%%%%

\subsection{Steps in observation campaign}
We imagine that the actual observation campaign consist of the following steps.
\begin{enumerate}[i]
%    \item \textbf{Perform spectroscopy on the reflected light from the target planet and determine its atmospheric composition. Retrain the neural network accordingly.}
    \item Determine the orbital elements of the Kepler orbit
    and the period from the (relative) positions of the planet at the observation points. \citet{Feng_2019} describe a comprehensive method to obtain these.
    \item Determine the diurnal period from light curves at (one or more) observation points, and the phase shift between the points \citep[see for example][]{Visser2015}.
    Close-in planets can be tidally locked or in a spin-orbit resonance. These need to be treated as special cases that we
    do not consider in this paper.
    \item
    Update the parameters, and correct 
    for the diurnal phase shift due to the light-travel (R{\o}mer) delay, at the different observation points.
    \item
    Retrieve the rotation axis as explained in Sect.\ \ref{sect_results_rotation_axis}.
    \item
    Retrieve the surface map. This is demonstrated in Sect.\ \ref{sect_surface_type_maps}.
\end{enumerate}

\subsection{Recommendations}
We have a number of recommendations for further research:
\begin{enumerate}[i]
\item Include more surface types in the training data, like ices 
      (such as those found on Earth, 
      Mars, and icy moons), differently colored deserts (such as 
      the martian red desert), and wetlands.
\item Implement dynamical surface maps to capture seasonal changes in e.g.\ 
      the size of polar ice caps, snow cover, and vegetation color, and train
      a network to recognize such surface types from their seasonal
      changes.
\item Include different model atmospheres and different cloud compositions, 
      such as sulfuric acid clouds as seen on Venus, that exhibit 
      angular features, such as the rainbow, at different phase 
      angles than water clouds
      \citep[see][for examples in single scattering phase functions]{scattering_in_atmospheres}.
\item Include other noise sources than photon noise, such as instrumental 
      noise and physical background noise sources.
\item Simulate with orbital and rotational periods that are closer together
      in frequency or in a resonance (like Mercury's $3:2$ spin-orbit
      resonance).
\item Account for signal smearing caused by the planet's rotation 
      during one observational integration period.
\item Optimize the number of sample points. Increasing the number of
      observational epochs may require new network architectures as the 
      number of trainable parameters may become too large for efficient
      training.
\item Test different weighting schemes for the loss function 
      of the neural network, in particular for surface type retrieval. 
      This would allow training to focus on facets with smaller or 
      or larger expected contributions to the phase curves.
\item Test the retrieval algorithms with light curves of Earth observed 
      from afar, e.g.\ using an instrument like LOUPE 
      \citep{loupe2} on the Moon or on a distant spacecraft.
\item Investigate how to adapt the network to identify 
      a surface type with an unknown albedo spectrum, such as that of
      vegetation that uses photosynthesis-like processes at other
      wavelengths than the terrestrial vegetation and/or that 
      has a red edge feature at different wavelengths.
\end{enumerate}

%--------------------------------------------------------------------------

%%%%%%%%%%%%%%%%%%%%%%%%%%%%%%%%%%%%%%%%%%%%%%%%%%%%%%%%%%%%%%%%%%%%%%%%%%%%%%
\section{Conclusions}
\label{sect_conclusions}

We used neural networks to retrieve surface and cloud maps of spatially 
unresolved Earth-like exoplanets, using temporal and spectral variations 
in the flux and polarization of the star light that is reflected by 
such planets. 

We were able to retrieve the rotation axes of our model planets using a 
neural network with 1-D convolutions. The mean squared error (MSE) is as 
small as $0.0097$ and depends on the orbital inclination angle.
In the special case of an exact edge-on orbit, the signals are insensitive 
to mirror reflection of the axis in the orbital plane. 
Up to this reflection symmetry, the rotation axis can be determined.

We have tested a new approach for planet mapping by predicting surface types
rather than surface albedos. A neural network with 1-D convolutions and
spherical convolutions applied in a ResNet fashion can create detailed 
planet maps that correctly predict up to 92\% of ocean facets and 
85\% of cloud, vegetation and desert facets on a planet. 
When applied to a model Earth, the retrieved map shows that the Sahara, 
Europe, Asia, the Americas and cloud patterns can all be retrieved. 
When photon noise is the dominant noise source, our results show that 
it should be possible to retrieve a map with such Earth-like patterns from 
future observations by the HabEx telescope in combination with a star shade
for planets at distances of up to $75~\text{ly}$.

We showed that the retrieval of the rotation axis and surface map is 
usually poor if the network assumes Lambertian reflection for 
a bidirectionally reflecting planet, except for planets in face-on orbits, 
where the lack of phase angle variation hides the bidirectionality 
of the actual reflection. 
In particular, retrieving maps of planets with oceanic reflection and 
atmospheric scattering while assuming Lambertian reflection results in
erroneous concentric bright patterns about the planet's poles due to 
an overestimation and an underestimation of the planet's brightness at 
small and large phase angles, respectively. 
The MSE of the rotation axis retrievals is reduced 
by a factor $10$ if the network does take bidirectional reflection 
into account. 

We also showed that while photon noise affects a planet's polarized 
signal more than its total flux signal, adding polarization
to the neural network training decreases the MSE of the 
rotation axis retrievals by about $15\%$ and increases the retrieval 
accuracy of ocean and and cloud-covered facets by $2\%$ and $1\%$, 
respectively, due to their unique polarization signatures.

In conclusion, neural networks appear to be promising tools for retrieving
exoplanet maps from total flux and polarization phase curves. 
Eight orbital locations and eight 
rotational phases appear to suffice for Earth-like planets 
when using the architectures we have discussed.

%%%%%%%%%%%%%%%%%%%%%%%%%%%%%%%%%%%%%%%%%%%%%%%%%%%%%%%%%%%%%%%%%%%%%%%%%%%%%%
\bibliographystyle{aa}
\bibliography{references}{}

\begin{thebibliography}{48}
\expandafter\ifx\csname natexlab\endcsname\relax\def\natexlab#1{#1}\fi

\bibitem[{Alsallakh {et~al.}(2021)Alsallakh, Kokhlikyan, Miglani, Yuan, \&
  Reblitz-Richardson}]{mind_the_pad}
Alsallakh, B., Kokhlikyan, N., Miglani, V., Yuan, J., \& Reblitz-Richardson, O.
  2021, in International Conference on Learning Representations

\bibitem[{{Asensio Ramos} \& {Pall{\'e}}(2021)}]{nn_cartography}
{Asensio Ramos}, A. \& {Pall{\'e}}, E. 2021, \aap, 646, A4

\bibitem[{Badshah {et~al.}(2017)Badshah, Ahmad, Rahim, \&
  Baik}]{speech_emotion_recognition}
Badshah, A., Ahmad, J., Rahim, N., \& Baik, S. 2017, in Speech Emotion
  Recognition from Spectrograms with Deep Convolutional Neural Network, 1--5

\bibitem[{{Bryson} {et~al.}(2020){Bryson}, {Coughlin}, {Batalha}, {Berger},
  {Huber}, {Burke}, {Dotson}, \& {Mullally}}]{2020AJ....159..279B}
{Bryson}, S., {Coughlin}, J., {Batalha}, N.~M., {et~al.} 2020, \apj, 159, 279

\bibitem[{{Cash}(2006)}]{star_shade_cash}
{Cash}, W. 2006, \nat, 442, 51

\bibitem[{{Dressing} \& {Charbonneau}(2015)}]{occurence_of_earths}
{Dressing}, C.~D. \& {Charbonneau}, D. 2015, \apj, 807, 45

\bibitem[{{Fan} {et~al.}(2019){Fan}, {Li}, {Li}, {Bartlett}, {Jiang}, {Natraj},
  {Crisp}, \& {Yung}}]{2D_alien_map}
{Fan}, S., {Li}, C., {Li}, J.-Z., {et~al.} 2019, \apjl, 882, L1

\bibitem[{{Farr} {et~al.}(2018){Farr}, {Farr}, {Cowan}, {Haggard}, \&
  {Robinson}}]{exocartographer}
{Farr}, B., {Farr}, W.~M., {Cowan}, N.~B., {Haggard}, H.~M., \& {Robinson}, T.
  2018, \aj, 156, 146

\bibitem[{Feng {et~al.}(2019)Feng, Lisogorskyi, Jones, Kopeikin, Butler,
  Anglada-Escud{\'{e}}, \& Boss}]{Feng_2019}
Feng, F., Lisogorskyi, M., Jones, H. R.~A., {et~al.} 2019, The Astrophysical
  Journal Supplement Series, 244, 39

\bibitem[{{Ford} {et~al.}(2001){Ford}, {Seager}, \&
  {Turner}}]{2001Natur.412..885F}
{Ford}, E.~B., {Seager}, S., \& {Turner}, E.~L. 2001, \nat, 412, 885

\bibitem[{{Fujii} \& {Kawahara}(2012)}]{fujii_2012}
{Fujii}, Y. \& {Kawahara}, H. 2012, \apj, 755, 101

\bibitem[{Gonz\'alez(2010)}]{fibonacci}
Gonz\'alez, A. 2010, Mathematical Geosciences, 42, 49

\bibitem[{{G{\'o}rski} {et~al.}(2005){G{\'o}rski}, {Hivon}, {Banday},
  {Wandelt}, {Hansen}, {Reinecke}, \& {Bartelmann}}]{gorski_2005}
{G{\'o}rski}, K.~M., {Hivon}, E., {Banday}, A.~J., {et~al.} 2005, \apj, 622,
  759

\bibitem[{{Groot} {et~al.}(2020){Groot}, {Rossi}, {Trees}, {Cheung}, \&
  {Stam}}]{colors_of_earth}
{Groot}, A., {Rossi}, L., {Trees}, V.~J.~H., {Cheung}, J.~C.~Y., \& {Stam},
  D.~M. 2020, \aap, 640, A121

\bibitem[{{Hansen} \& {Travis}(1974)}]{scattering_in_atmospheres}
{Hansen}, J.~E. \& {Travis}, L.~D. 1974, \ssr, 16, 527

\bibitem[{He {et~al.}(2015)He, Zhang, Ren, \& Sun}]{prelu}
He, K., Zhang, X., Ren, S., \& Sun, J. 2015, IEEE International Conference on
  Computer Vision (ICCV 2015), 1502

\bibitem[{He {et~al.}(2016)He, Zhang, Ren, \& Sun}]{resnet}
He, K., Zhang, X., Ren, S., \& Sun, J. 2016, in 2016 IEEE Conference on
  Computer Vision and Pattern Recognition (CVPR), 770--778

\bibitem[{Hornik(1991)}]{HORNIK1991251}
Hornik, K. 1991, Neural Networks, 4, 251

\bibitem[{{Hunziker} {et~al.}(2020){Hunziker}, {Schmid}, {Mouillet}, {Milli},
  {Zurlo}, {Delorme}, {Abe}, {Avenhaus}, {Baruffolo}, {Bazzon}, {Boccaletti},
  {Baudoz}, {Beuzit}, {Carbillet}, {Chauvin}, {Claudi}, {Costille}, {Daban},
  {Desidera}, {Dohlen}, {Dominik}, {Downing}, {Engler}, {Feldt}, {Fusco},
  {Ginski}, {Gisler}, {Girard}, {Gratton}, {Henning}, {Hubin}, {Kasper},
  {Keller}, {Langlois}, {Lagadec}, {Martinez}, {Maire}, {Menard}, {Meyer},
  {Pavlov}, {Pragt}, {Puget}, {Quanz}, {Rickman}, {Roelfsema}, {Salasnich},
  {Sauvage}, {Siebenmorgen}, {Sissa}, {Snik}, {Suarez}, {Szul{\'a}gyi},
  {Thalmann}, {Turatto}, {Udry}, {van Holstein}, {Vigan}, \& {Wildi}}]{sphere}
{Hunziker}, S., {Schmid}, H.~M., {Mouillet}, D., {et~al.} 2020, \aap, 634, A69

\bibitem[{JPL(2019)}]{hab_ex}
JPL. 2019, HabEx - Habitable Exoplanet Observatory Final Report, Tech. rep.,
  NASA

\bibitem[{{Karalidi} {et~al.}(2011){Karalidi}, {Stam}, \&
  {Hovenier}}]{polarization_of_clouds}
{Karalidi}, T., {Stam}, D.~M., \& {Hovenier}, J.~W. 2011, \aap, 530, A69

\bibitem[{{Karalidi} {et~al.}(2012){Karalidi}, {Stam}, \&
  {Hovenier}}]{2012A&A...548A..90K}
{Karalidi}, T., {Stam}, D.~M., \& {Hovenier}, J.~W. 2012, \aap, 548, A90

\bibitem[{Karras {et~al.}(2020)Karras, Laine, Aittala, Hellsten, Lehtinen, \&
  Aila}]{stylegan2}
Karras, T., Laine, S., Aittala, M., {et~al.} 2020, in Analyzing and Improving
  the Image Quality of StyleGAN, 8107--8116

\bibitem[{{Kawahara}(2016)}]{frequency_modulation}
{Kawahara}, H. 2016, \apj, 822, 112

\bibitem[{{Kawahara} \& {Masuda}(2020)}]{sot_dynamic}
{Kawahara}, H. \& {Masuda}, K. 2020, \apj, 900, 48

\bibitem[{{Kingma} \& {Ba}(2014)}]{adam}
{Kingma}, D.~P. \& {Ba}, J. 2014, arXiv e-prints, arXiv:1412.6980

\bibitem[{{Kiranyaz} {et~al.}(2021){Kiranyaz}, {Avci}, {Abdeljaber}, {Ince},
  {Gabbouj}, \& {Inman}}]{1D_convolutions}
{Kiranyaz}, S., {Avci}, O., {Abdeljaber}, O., {et~al.} 2021, Mechanical Systems
  and Signal Processing, 151, 107398

\bibitem[{{Klind{\v{z}}i{\'c}} {et~al.}(2021){Klind{\v{z}}i{\'c}}, {Stam},
  {Snik}, {Keller}, {Hoeijmakers}, {van Dam}, {Willebrands}, {Karalidi},
  {Pallichadath}, {van Dijk}, \& {Esposito}}]{loupe2}
{Klind{\v{z}}i{\'c}}, D., {Stam}, D.~M., {Snik}, F., {et~al.} 2021,
  Philosophical Transactions of the Royal Society of London Series A, 379,
  20190577

\bibitem[{Krachmalnicoff \& Tomasi(2019)}]{spherical_convolutions}
Krachmalnicoff, N. \& Tomasi, M. 2019, \aap, 628

\bibitem[{Lecun {et~al.}(1989)Lecun, Boser, Denker, Henderson, Howard, Hubbard,
  \& Jackel}]{lecun1989}
Lecun, Y., Boser, B., Denker, J., {et~al.} 1989, Neural Information Processing
  Systems, 2, 396

\bibitem[{{Maier} {et~al.}(2021){Maier}, {Zellem}, {Colavita}, {Mennesson},
  {Bailey}, {Nemati}, {Ygouf}, \& {Douglas}}]{2021AAS...23732703M}
{Maier}, E.~R., {Zellem}, R.~T., {Colavita}, M., {et~al.} 2021, in American
  Astronomical Society Meeting Abstracts, Vol.~53, American Astronomical
  Society Meeting Abstracts, 327.03

\bibitem[{{Mayor} \& {Queloz}(1995)}]{first_exoplanet}
{Mayor}, M. \& {Queloz}, D. 1995, \nat, 378, 355

\bibitem[{Mogensen(2010)}]{map_generation}
Mogensen, T.~{\AE}. 2010, in Perspectives of Systems Informatics, ed.
  A.~Pnueli, I.~Virbitskaite, \& A.~Voronkov (Berlin, Heidelberg: Springer
  Berlin Heidelberg), 306--318

\bibitem[{{National Academies of Sciences, Engineering, and
  Medicine}(2021)}]{NAP26141}
{National Academies of Sciences, Engineering, and Medicine}. 2021, Pathways to
  Discovery in Astronomy and Astrophysics for the 2020s (Washington, DC: The
  National Academies Press)

\bibitem[{{Rossi} {et~al.}(2018){Rossi}, {Berzosa-Molina}, \&
  {Stam}}]{pymiedap}
{Rossi}, L., {Berzosa-Molina}, J., \& {Stam}, D.~M. 2018, \aap, 616, A147

\bibitem[{Schmidhuber(2015)}]{nn_overview}
Schmidhuber, J. 2015, Neural Networks, 61, 85

\bibitem[{Seager \& Kasdin(2018)}]{roman_starshade}
Seager, S. \& Kasdin, N. 2018, Starshade Rendezvous Probe Study Report, Nasa
  astrophysics probe study, JPL, Goddard Space Flight Center, MIT, Princeton
  University, Northrup Grumman

\bibitem[{{Seager} {et~al.}(2005){Seager}, {Turner}, {Schafer}, \&
  {Ford}}]{vegetation_biosignatures}
{Seager}, S., {Turner}, E.~L., {Schafer}, J., \& {Ford}, E.~B. 2005,
  Astrobiology, 5, 372

\bibitem[{{Snellen} {et~al.}(2021){Snellen}, {Snik}, {Kenworthy}, {Albrecht},
  {Anglada-Escud{\'e}}, {Baraffe}, {Baudoz}, {Benz}, {Beuzit}, {Biller},
  {Birkby}, {Boccaletti}, {van Boekel}, {de Boer}, {Brogi}, {Buchhave},
  {Carone}, {Claire}, {Claudi}, {Demory}, {D{\'e}sert}, {Desidera}, {Gaudi},
  {Gratton}, {Gillon}, {Grenfell}, {Guyon}, {Henning}, {Hinkley}, {Huby},
  {Janson}, {Helling}, {Heng}, {Kasper}, {Keller}, {Krause}, {Kreidberg},
  {Madhusudhan}, {Lagrange}, {Launhardt}, {Lenton}, {Lopez-Puertas}, {Maire},
  {Mayne}, {Meadows}, {Mennesson}, {Micela}, {Miguel}, {Milli}, {Min}, {de
  Mooij}, {Mouillet}, {N'Diaye}, {D'Orazi}, {Palle}, {Pagano}, {Piotto},
  {Queloz}, {Rauer}, {Ribas}, {Ruane}, {Selsis}, {Sozzetti}, {Stam}, {Stark},
  {Vigan}, \& {de Visser}}]{2021ExA...tmp..124S}
{Snellen}, I. A.~G., {Snik}, F., {Kenworthy}, M., {et~al.} 2021, Experimental
  Astronomy

\bibitem[{{Stam}(2008)}]{stam_2008}
{Stam}, D.~M. 2008, \aap, 482, 989

\bibitem[{{Stam} {et~al.}(2006){Stam}, {de Rooij}, {Cornet}, \&
  {Hovenier}}]{stam_2006}
{Stam}, D.~M., {de Rooij}, W.~A., {Cornet}, G., \& {Hovenier}, J.~W. 2006,
  \aap, 452, 669

\bibitem[{{Stam} {et~al.}(2004){Stam}, {Hovenier}, \&
  {Waters}}]{2004A&A...428..663S}
{Stam}, D.~M., {Hovenier}, J.~W., \& {Waters}, L.~B.~F.~M. 2004, \aap, 428, 663

\bibitem[{{Trees} \& {Stam}(2019)}]{blue_white_red}
{Trees}, V.~J.~H. \& {Stam}, D.~M. 2019, \aap, 626, A129

\bibitem[{{Tuomi} {et~al.}(2014){Tuomi}, {Jones}, {Barnes},
  {Anglada-Escud{\'e}}, \& {Jenkins}}]{all_stars_have_planets_1}
{Tuomi}, M., {Jones}, H. R.~A., {Barnes}, J.~R., {Anglada-Escud{\'e}}, G., \&
  {Jenkins}, J.~S. 2014, \mnras, 441, 1545

\bibitem[{{van de Hulst}(1957)}]{1957lssp.book.....V}
{van de Hulst}, H.~C. 1957, {Light Scattering by Small Particles} ({New York \&
  London}: {John Wiley and Sons \& Chapman and Hall})

\bibitem[{{Visser} \& {van de Bult}(2015)}]{Visser2015}
{Visser}, P.~M. \& {van de Bult}, F.~J. 2015, \aap, 579, A21

\bibitem[{Wang {et~al.}(2018)Wang, Chen, Yuan, Liu, Huang, Hou, \&
  Cottrell}]{semantic_segmentation}
Wang, P., Chen, P., Yuan, Y., {et~al.} 2018, in 2018 IEEE Winter Conference on
  Applications of Computer Vision (WACV), 1451--1460

\bibitem[{Zhang \& Sabuncu(2018)}]{categorical_cross_entropy}
Zhang, Z. \& Sabuncu, M. 2018, in Advances in Neural Information Processing
  Systems, ed. S.~Bengio, H.~Wallach, H.~Larochelle, K.~Grauman,
  N.~Cesa-Bianchi, \& R.~Garnett, Vol.~31 (Curran Associates, Inc.)

\end{thebibliography}

%%%%%%%%%%%%%%%%%%%%%%%%%%%%%%%%%%%%%%%%%%%%%%%%%%%%%%%%%%%%%%%%%%%%%%%%%%%%%%
\end{document}